\documentclass[preprint,12pt]{elsarticle}
\usepackage{geometry}
\usepackage{amssymb}
\usepackage{amsmath}
\usepackage{multirow}%
\usepackage{amsmath,amssymb,amsfonts}%
\usepackage{amsthm}%
\usepackage{mathrsfs}%
\usepackage[title]{appendix}%
\usepackage{xcolor}%
\usepackage{soul}
\usepackage[dvipsnames]{xcolor}
\usepackage{textcomp}%
\usepackage{manyfoot}%
\usepackage{booktabs}%
\usepackage{algorithm}%
\usepackage{algorithmicx}%
\usepackage{algpseudocode}%
\usepackage{listings}%
\usepackage{url,hyperref,lineno,microtype}
\usepackage[onehalfspacing]{setspace}
\usepackage{array}
\usepackage{graphicx} % for including images
\usepackage{caption}  % optional, improves captions
\usepackage{wrapfig}  % for wrapping text around figures
\usepackage[rightcaption]{sidecap} % in preamble
\usepackage{multicol} % in your preamble
\usepackage{longtable}
\newcommand{\bv}{\mathbf{v}}
\newcommand{\bn}{\mathbf{n}}

\newcommand{\bvg}{\hat{\mathbf{v}}}

\newcommand{\bs}[1]{\boldsymbol{#1}}

\newcommand{\greentxt}[1]{\textcolor{black}{#1}}

\usepackage[table]{xcolor}
\definecolor{datagray}{RGB}{235,235,235}
\definecolor{resultbeige}{RGB}{242,236,205}
\definecolor{err0}{RGB}{248,248,240}
\definecolor{err5}{RGB}{235,230,190}
\definecolor{err25}{RGB}{232,220,140}
\definecolor{err50}{RGB}{238,214,80}
\definecolor{err100}{RGB}{248,190,0}

\journal{Computers in Biology and Medicine}

\begin{document}

\begin{frontmatter}

\title{Towards Modeling the Hemodynamic Impact of Mitral and Aortic Valve Repair in Patients with Left Ventricular Assist Devices}

\author[label1]{Mia Bonini\corref{cor1}}\ead{mbonini@umich.edu}
\author[label1]{Michael Ferguson}%\ead{michaef@umich.edu}
\author[label2]{Marc Hirshvogel}%\ead{marc.hirschvogel@ambit.net}
\author[label3]{Maximilian Balmus}%\ead{mbalmus@turing.ac.uk}
\author[label4]{Paul C. Tang}%\ead{Tang.Paul2@mayo.edu}
\author[label5]{Francis Pagani}%\ead{fpagani@med.umich.edu}
\author[label1,label5]{David Nordsletten}\ead{nordslet@umich.edu}

\affiliation[label1]{organization = {Department of Biomedical Engineering, University of Michigan},
            city = {Ann Arbor},
            state = {MI},
            country = {USA}}
\affiliation[label2]{organization = {Division 2.2 Process Simulation, Bundesanstalt für Materialforschung und -prüfung (BAM)},
            city = {Berlin},
            country = {Germany}}
\affiliation[label3]{organization = {The Alan Turing Institute},
            city = {London},
            country = {United Kingdon}}
\affiliation[label4]{organization = {Department of Cardiac Surgery, Mayo Clinic},
            city = {Rochester},
            state = {MN},
            country = {USA}}
\affiliation[label5]{organization = {Department of Cardiac Surgery, University of Michigan},
            city = {Ann Arbor},
            state = {MI},
            country = {USA}}

\cortext[cor1]{Corresponding author}

%% Abstract
\begin{abstract}
Valve dysfunction is a major threat to long-term success in left ventricle assist device (LVAD) therapy, with direct implications for right heart performance. In this study, we apply a patient-specific, image-based computational modeling framework to evaluate the hemodynamic impact of simulated mitral and aortic valve repair in five LVAD-supported patients. Each patient was modeled under four conditions: 
(patient-specific LVAD-supported state), simulated mitral valve (MV) repair, simulated aortic valve (AV) repair, and simulated combined MV\&AV repair. Because validation data for valve repair were unavailable, the simulated repair scenarios are exploratory \textit{in silico} interventions based on clinically validated patient-specific models.
The models integrate dynamic CT imaging, echocardiography, catheterization data, and device-specific LVAD parameters into a coupled 3D–0D simulation pipeline. Valve dynamics are governed by transvalvular pressure and flow, allowing physiological modeling of regurgitant lesions and surgical repair. The right ventricular (RV) was assessed using a combination of model-derived metrics, including right ventricular ejection fraction (RVEF), pulmonary artery pulsatility index (PAPi), and RV-PA coupling. In addition, we performed blood residence time (RT) analysis to evaluate blood stasis within the left heart and aorta.
The simulations suggest that valve repair improved cardiac output, reduced pulmonary congestion, and enhanced right ventricular loading conditions. Notably, mitral valve repair restored aortic valve opening during the cardiac cycle, which improved sinus washout and reduced blood residence time within the aortic root—factors associated with lower thrombotic risk. 
Overall, these findings suggest a potential role for valve repair in LVAD-supported hearts, though larger, validated patient cohorts are needed to confirm these results.
\end{abstract}


\begin{highlights}
\item Patient-specific 3D–0D hemodynamic models simulating aortic and mitral valve repair effects in LVAD hearts
\item simulated mitral repair improves aortic valve opening under LVAD support
\item simulated aortic and mitral valve repair improves right ventricle loading conditions
\item Residence time analysis reveals reduced aortic root stasis after simulated mitral valve repair
\end{highlights}

%% Keywords

\end{frontmatter}



% ---------------------
% ---------------------
\section{Introduction} 

Heart failure remains a leading cause of morbidity and mortality worldwide, and left ventricular assist devices (LVAD) are increasingly used to provide circulatory support in patients with advanced disease \cite{Roth2018,Woodruff2024,Yuzefpolskaya2023}.
By generating continuous-flow support, LVADs unload the left ventricle and improve survival and quality of life. At the same time, they fundamentally alter cardiac physiology—impacting chamber filling, valve dynamics, and myocardial loading—in a population already characterized by advanced, complex disease \cite{Jezovik2017}.
Within this altered physiological state, patients may experience right heart failure (RHF), persistent mitral regurgitation (MR), or develop de novo aortic insufficiency (AI) during device support \cite{Fatullayev2015,Bravo2022,Cowger2010,Kohno2025}.
The role of valve repair in this population is debated.
Clinical studies report mixed outcomes, with some showing reductions in pulmonary pressures and improved symptoms after MR repair, while others find no survival benefit with increased surgical risk \cite{Mehra2022-mj,Kherallah2024,Kanwar2020,Rad2023,Noly2022}.
These inconsistencies likely stem from patient heterogeneity, differences in surgical technique, and the reliance on outcome-based statistics rather than direct assessment of cardiovascular mechanics.
Thus, it remains unclear whether repairing left-sided valves in LVAD patients provides hemodynamic benefit, and if so, through which mechanisms.

Computational modeling helps address this gap by enabling controlled, image-based studies of valve dysfunction and repair while isolating patient-specific variables. It also allows quantification of hemodynamic metrics like residence time -- a surrogate for blood stasis and thrombosis risk in LVAD patients -- which can be difficult to evaluate clinically~\cite{Sahni2023,Esmaily-Moghadam2013,McCormick2014}. Residence time may be estimated using a variety of approaches, including transport analysis based on Lagrangian coherent structures derived from imaging data~\cite{Hendabadi2013-xj}, although in the present study it is computed directly from the simulated flow field. While prior CFD studies have examined blood stasis, they often focus on isolated regions and do not capture whole-heart interactions or how valve pathology alters flow exchange, residence time, and ventricular loading under LVAD support.
Several computational studies have investigated LVAD physiology and valve dysfunction, providing valuable frameworks for modeling circulatory dynamics in this population \cite{McCormick2011,Jelenc2013,Kim2018}. 
Lumped-parameter (0D) models have also been used to study the effects of mitral regurgitation on right ventricular afterload under LVAD support, but an open-loop model and lack of 3D flow limits the assessment of global hemodynamics and intraventricular flow features ~\cite{Jelenc2013}.
Other works have coupled 3D electromechanical–circulatory model to examine MR and AI effects on ventricular mechanics and LVAD performance; however, its focus on myocardial mechanics restricted detailed analysis of fluid dynamics such as flow stasis and residence time~\cite{Kim2018}.
Building on these contributions, a fully three-dimensional, flow-resolved modeling approach would enable the quantification of valve failure and the influence of subsequent repair on intracardiac flow distribution, blood stasis, and ventricular loading under controlled conditions.

In this work, we developed patient-specific models for five LVAD patients to evaluate how mitral and aortic valve repair modify left heart hemodynamics. 
The model provides a fully time-resolved, 3D representation of left heart flow under LVAD support, including valve opening and closing, ventricular unloading, pump function, and systemic coupling \cite{Bonini2025b}. 
By performing controlled \textit{in silico} interventions in the same patients, we isolate how valve repair may influence LVAD physiology, quantifying changes in cardiac output, aortic valve opening, blood stasis, and right ventricle clinical metrics. These simulations are intended to provide \textit{in silico} evaluation of the hemodynamic consequences of valve repair, rather than validated reconstructions of postoperative outcomes.

% ---------------------
% ---------------------
\section{Methods} \label{methods}
To capture heterogeneity in the patient population, we model five patient-specific LVAD patients using methods we developed in \cite{Bonini2025b}. 
From there, we repair the AV, MV, and AV\&MV for each patient. The following sections outline the clinical data utilized (Sec. \ref{sec:data}), the construction of the three-dimensional cardiac model (Sec. \ref{sec:segmentation}), and the coupled zero-dimensional representation of the systemic and pulmonary circulation (Sec. \ref{sec:0Dfitting}).
Subsequent sections describe the governing equations for the CFD and advection–diffusion formulations (Sec. \ref{sec:cfd}), the implementation of valve repair (Sec. \ref{sec:vlvrepair}), and the quantitative metrics analyzed in this study (Sec.~\ref{sec:quant}). 

% ---------------------
\subsection{Patient Data} \label{sec:data}
For each patient in the LVAD cohort, post-LVAD multimodal clinical data was used to develop the models, including: dynamic contrast-enhanced computed tomography (CT), non-invasive blood pressure measurements, catheter-based hemodynamics, and echocardiographic assessments.
CT scans were performed on a 64-slice helical scanner (Siemens Somatom Force) with in-plane resolutions of 0.4886 mm (sagittal) and 0.4885 mm (coronal), and a through-plane resolution of 1.25 mm (axial). 
Image acquisition was contrast-enhanced with iopamidol, and retrospective ECG gating was used to reconstruct 10-20 cardiac phases.
Device-specific information, including LVAD type (HeartMate 3, Abbott Laboratories, Chicago, IL), operating speed, and pump flow rate, was incorporated for each patient to accurately represent left heart and aortic hemodynamics under mechanical support. A summary of data used for each patient is shown in Table \ref{tbl:data}.
All imaging and hemodynamic data collection were performed under IRB-approved protocol HUM00196629 (approved April 2021).
 
% ADD PARAGRAPH 
The five patients considered in this study exhibit substantial heterogeneity in both cardiac structure and hemodynamic state (see  Table \ref{tbl:data}). Ventricular volumes vary markedly across cases, with end-diastolic volume indices (EDVI) ranging from 77.8 to 313.5 mL/m$^2$ and end-systolic volume indices (ESVI) ranging from 64 to 295 mL/m$^2$, resulting in stroke volume indices (SVI) spanning 11.5 to 14.4 mL/m$^2$. These differences reflect varying degrees of ventricular dilation and contractile function under LVAD support. 
Prior to LVAD implantation, all patients had at least moderate MR. \greentxt{After the LVAD procedure the presence and severity of mitral regurgitation ranged from mild to moderate (Table \ref{tbl:data}). The patients also had aortic insufficiency ranging from none to moderate/severe.} The severity of regurgitation and insufficiency was determined based on an echocardiogram (TTE) exam.
Variability is also observed in global hemodynamic measures, including cardiac output, pulmonary pressures, and right ventricular loading conditions. 
Beyond hemodynamics, the cohort includes patients with differing demographics, with heights ranging from 63 to 74 in, weights from 130 to 243 lb, and representation of both male and female patients. 

\begin{table}[!htb]
\small
\renewcommand{\arraystretch}{1.1} % Increases row height by 1.5x
\centering
\begin{tabular}{ | >{\centering\arraybackslash} m{5.0cm}| >{\centering\arraybackslash}m{1.95cm}|>{\centering\arraybackslash}m{1.95cm}|>{\centering\arraybackslash}m{1.95cm}|>{\centering\arraybackslash}m{1.95cm}|>{\centering\arraybackslash}m{1.95cm}|}
 \hline
\textbf{Data} & \textbf{Pat. A} & \textbf{Pat. B} & \textbf{Pat. C} & \textbf{Pat. D} & \textbf{Pat. E} \\ \hline
\textit{Sex} & Female & Male & Male & Male & Male \\ \hline
\textit{Height (inch)} & 63 & 72 & 67 & 74 & 68 \\ \hline
\textit{Weight (lb)} & 130 & 197 & 151 & 230 & 243 \\ \hline
\textit{BSA (m$^2$)} & 1.61 & 2.11 & 1.79 & 2.37 & 2.27  \\ \hline
\textit{LV ESVI (mL/m$^2$)} & 64.0 & 205.0 & 153.4 & 295.1 & 187.8 \\ \hline
\textit{LV EDVI (mL/m$^2$)} & 77.8 & 219.3 & 166.9 & 313.5 & 199.3  \\ \hline
\textit{LV Wall Volume (mL)} & 126.6 & 326.0 & 207.6 & 386.5 & 276.2 \\ \hline
\textit{Syst. sBP/dBP (mmHg)} & 94/68 & 106/83 & 95/61  & 116/81 & 94/65 \\ \hline
\textit{mean }\& \textit{max RAP (mmHg)} & 5 \& 8 & 3 \& 4 & 19 \& 23 & 5 \& 7 & 26 \& 20 \\ \hline
\textit{RV sBP/dBP (mmHg)} & 20/3 & 29/4 & 26/13 & 32/4 & 48/14 \\ \hline
\textit{RVEDP (mmHg)} & 3 & 5 & 16 & 7 & 20 \\ \hline
\textit{mean }\& \textit{max PCWP (mmHg)} &  12 \& 13 & 11 \& 10 & 15 \& 18 & 10 \& 11 & 25 \& 21 \\ \hline
\textit{sPAP/dPAP (mmHg)} & 21/12 & 31/18 & 27/20 & 31/15 & 44/30 \\ \hline
\textit{mean PAP (mmHg)} & 15 & 24 & 22 & 21 & 35 \\ \hline
\textit{RVEDV (mL)} & 185 & 234 & 316 & 395	& 323 \\ \hline
\textit{RA Vol. (mL)} & 154 & 151 & 288 & 161 & 172 \\ \hline
\textit{LVAD Output (L/min)} & 3.7 & 5.1 & 4.0 & 4.5 & 3.4 \\ \hline
\textit{Coronary Output (L/min)} & 0.13 & 0.28 & 0.18 & 0.33 & 0.23 \\ \hline
\textit{MV Regurgitation} & mild & mod. & mild & mild & mild \\ \hline
\textit{AV Insufficiency} & mild & mod./sev. & mild/mod. & mild/mod. & none \\ \hline
\textit{TV Regurgitation} & mod. & none & mild & mild & mild \\ \hline
\textit{PV Insufficiency} & mild & none & none & mild & none \\ \hline
\textit{Pump Speed (rpm)} & 5400 & 5900 & 5200 & 5700 & 6000 \\  \hline
% \textit{LVAD Device} & HeartMate 3 & HeartMate 3 & HeartMate 3 & HeartMate 3 & HeartMate 3  \\ \hline
\end{tabular}
\caption{post-LVAD patient data from echocardiogram, right heart catheterization, and LVAD device. Abbreviations: \textit{BSA}, body surface area; \textit{LV}, left ventricle; \textit{ESVI}, end systolic volume index; \textit{EDVI}, end diastolic volume index; Abbreviations: \textit{CO}, cardiac output; \textit{sBP}, systolic blood pressure; \textit{dBP}, diastolic blood pressure; \textit{RAP}, right atrial pressure; \textit{RVP}, right ventricle pressure; \textit{PAP}, pulmonary artery pressure;\textit{PCWP}, pulmonary capillary wedge pressure; \textit{RVEDP}, right ventricle end diastolic pressure; \textit{RF}, regurgitant fraction; \textit{RVEDV}, right ventricle end diastolic volume; \textit{Vol}, Volume; \textit{mod}, moderate; \textit{sev}, severe.}
\label{tbl:data}
\end{table}

% ---------------------
\subsection{Anatomic Reconstruction and 3D Model Generation} \label{sec:segmentation}

\begin{figure}[!htb]
\centering
\includegraphics[width = 0.8\textwidth]{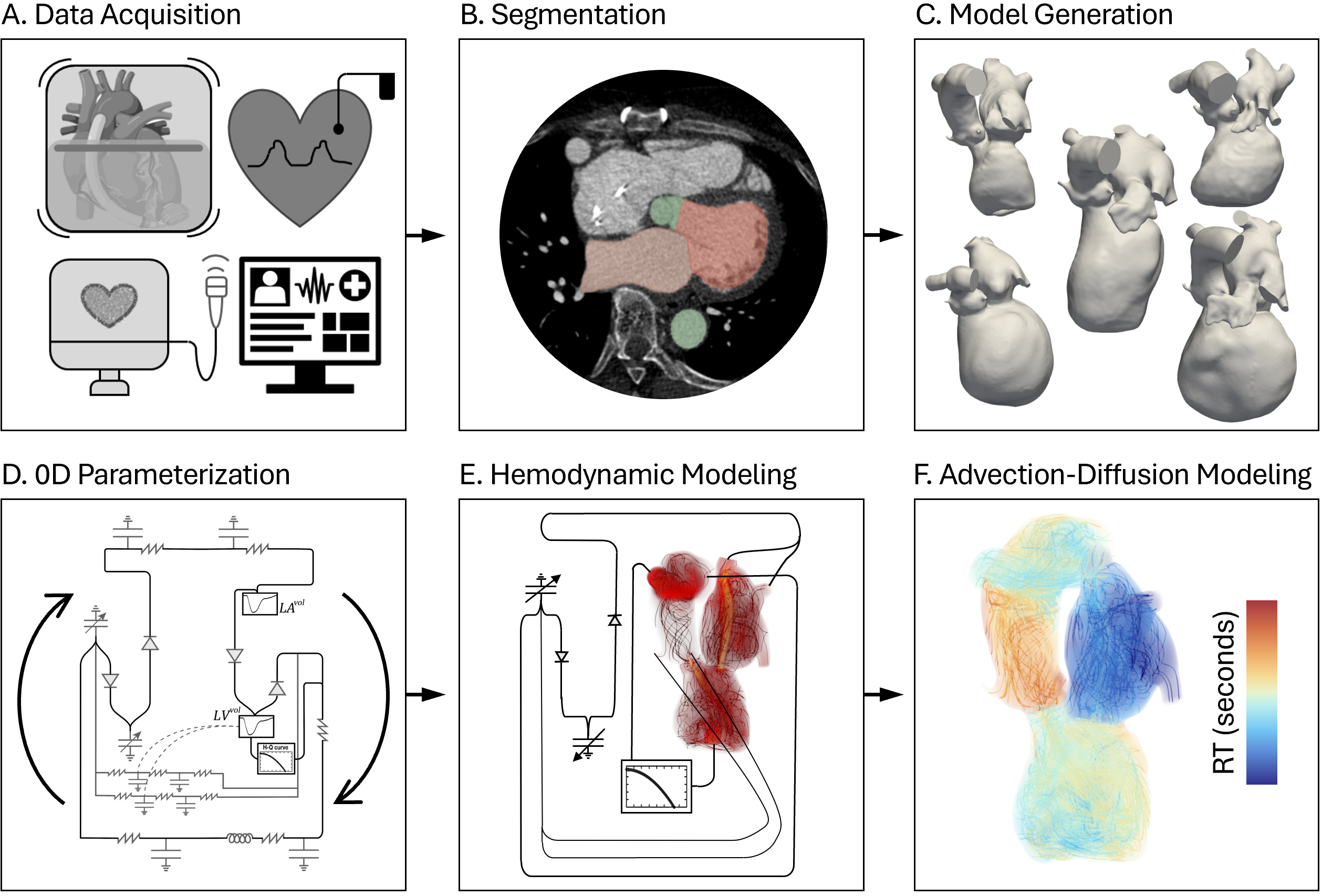}
\caption{Modeling Pipeline A) Data collection (dynamic CT imaging with contrast; right heart catheterization; echocardiogram; medical records), B) Segmentation (LV, LA, aorta, LVAD cannula), C) From the segmentation, the 5 patient models are generated. D) 0D parameter optimization using patient data, E) 3D-0D CFD modeling, F) Advection-Diffusion Modeling solving for residence time.} 
\label{fig:pipeline}
\end{figure}

Contrast-enhanced axial CT scans at end diastole were used to reconstruct patient-specific anatomy. 
The presence of metal-related artifacts from the implanted LVAD reduced the reliability of automated segmentation pipelines based on neural networks \cite{Xu2021}. To address this, blood pool geometries of the left ventricle (LV), left atrium (LA), aorta, coronary arteries, and both LVAD cannulae were delineated using a combination of semi-automatic and manual segmentation in 3D Slicer~\cite{slicer} (Fig. \ref{fig:pipeline}B). The right ventricle and right atrium were also segmented at end-diastole to calculate chamber volumes, which were used to estimate parameters in the 0D model. This is further explained in Section \ref{sec:0Dfitting}.
The LV myocardial wall was additionally segmented to compute myocardial volume, which was used in estimating coronary flow demand~\cite{Barral2011}. For all models, a tetrahedral volume mesh was generated using SimModeler meshing tools~\cite{2024SimModeler} with an average edge length of 1.0~mm. A mesh convergence study is reported in the Supplementary \ref{supp:mesh}.

\begin{figure}[!htb]
    \centering
    \includegraphics[width = \textwidth]{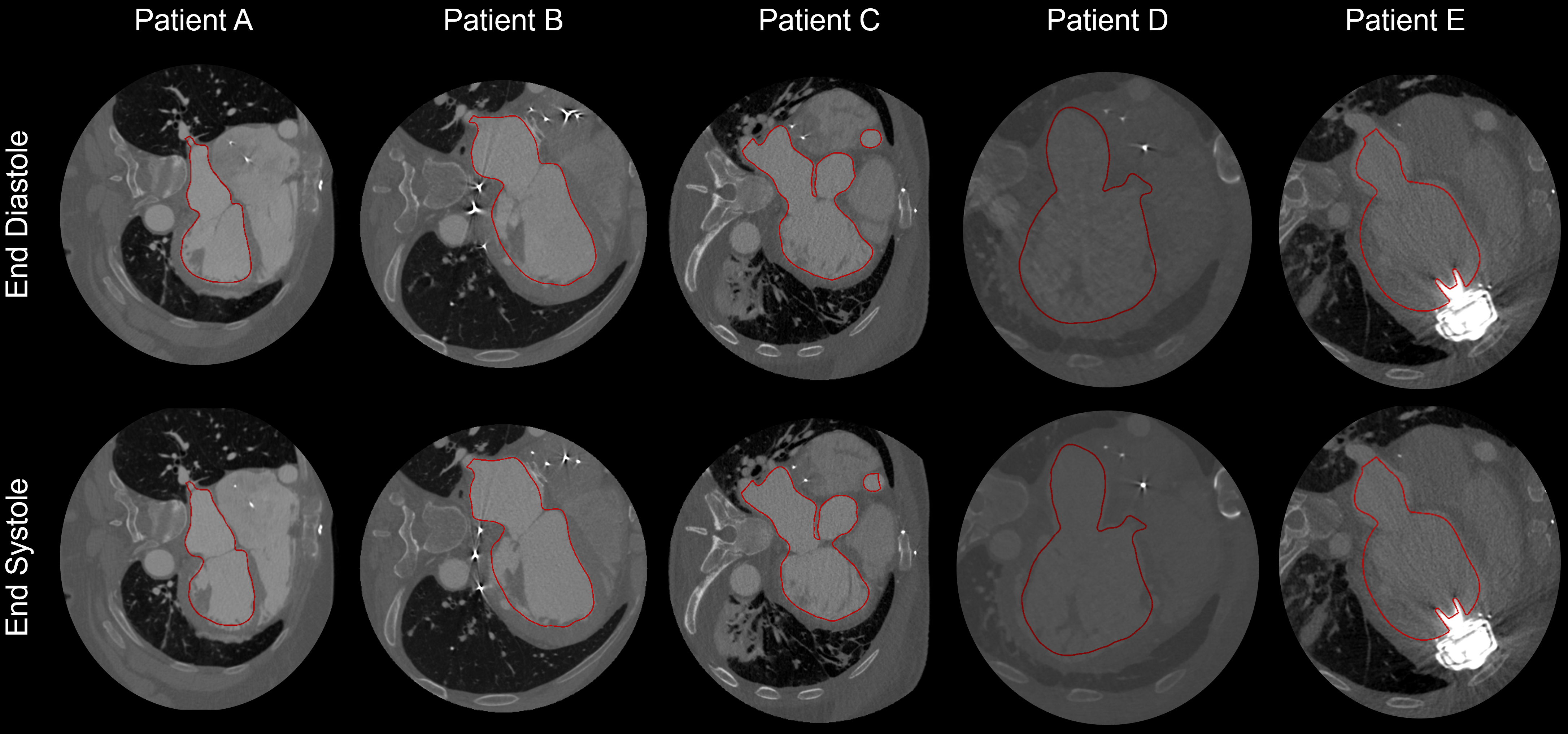}
    \caption{End diastolic and end systolic time phase showing the CT image with contrast and model edge in red.}
    \label{fig:track}
\end{figure}

To represent cardiac motion, non-rigid image registration was applied to the time-resolved CT data using the Image Registration Toolkit (IRTK)~\cite{Chandrashekara,Shi2012}. 
This provided deformation fields that mapped anatomical changes throughout the cardiac cycle (Fig. \ref{fig:track}). 
The resulting displacement information was transferred onto the segmented surfaces, producing a series of meshes corresponding to each CT phase. 
Temporal interpolation between reconstructed phases generated continuous boundary displacement data, which was then differentiated with respect to time to calculate domain boundary velocities ($\hat{\mathbf{v}}_D$) over the entire cardiac cycle. 

\begin{figure}[!htb]
    \centering
    \includegraphics[width = 0.8\textwidth]{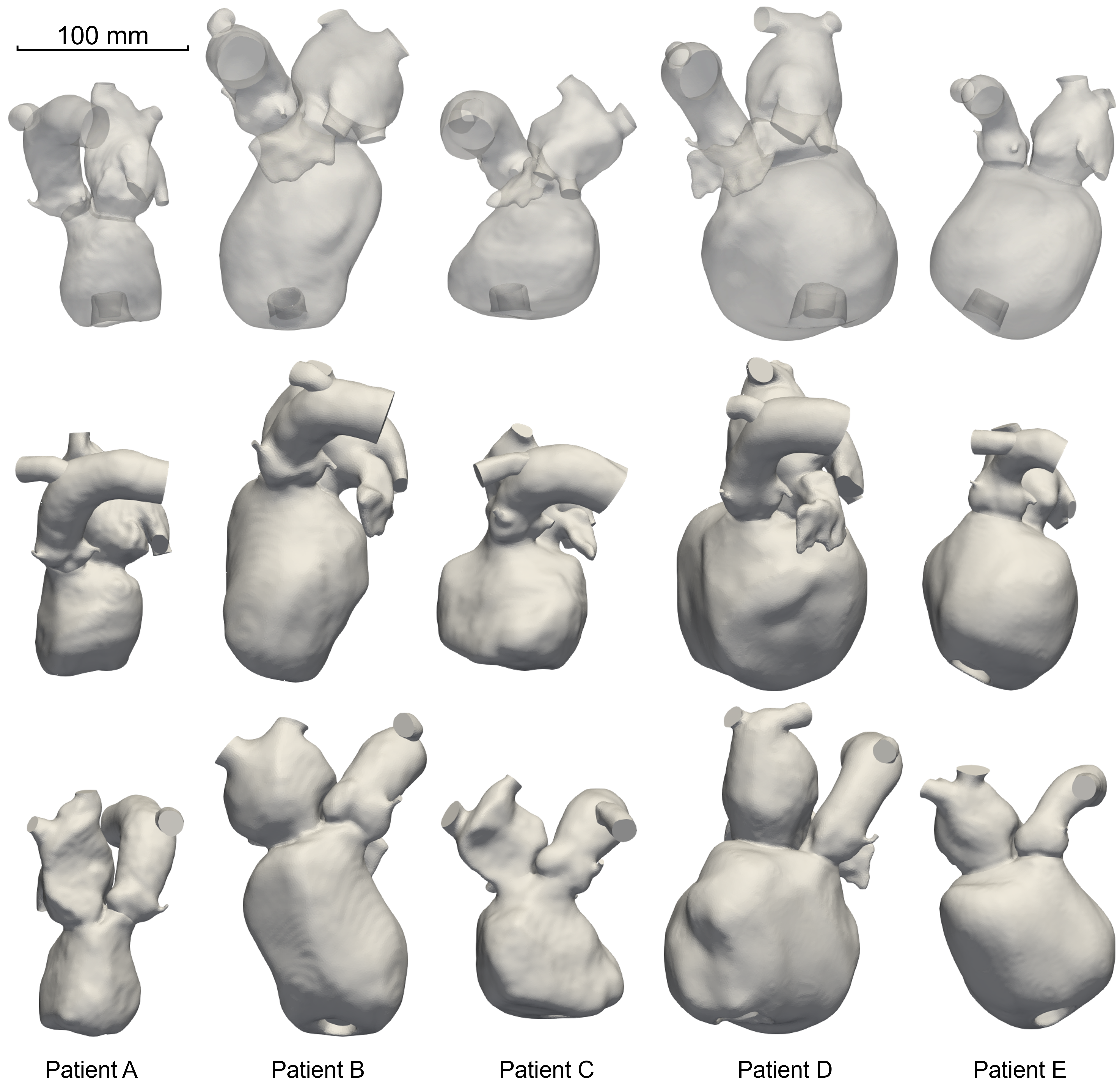}
    \caption{Three rotated views of the 5 patients (A--E) 3D heart models reconstructed at end-diastole. The first view has transparent faces to see the placement of the LVAD inflow cannula in the LV. Models are shown to scale using the 100 mm reference bar.}
    \label{fig:models}
\end{figure}

% ---------------------
\subsection{0D Model Parameterization} \label{sec:0Dfitting}
A lumped-parameter (0D) model was implemented to represent the global circulation, including the cardiac chambers, systemic, pulmonary, and coronary vasculature, together with the LVAD outflow circuit (Fig. \ref{fig:pipeline}D). The framework was based on the multi-compartment model developed by Hirschvogel et al.~\cite{Hirschvogel2017}, consisting of 29 state variables and 23 adjustable parameters. Vascular compartments were described using resistance–inductance–capacitance (RLC) analogues, the right heart was modeled with time-varying elastance functions, and patient-specific LA and LV volumes from imaging were prescribed to align the 0D and 3D domains. Regurgitant valve lesions were included by assigning effective regurgitant orifice areas constrained by clinical severity~\cite{Franz2021}. Clinical severity was determined from each patient's post-LVAD echocardiogram report.

Personalization of the 0D model was achieved by adjusting all parameters to match available clinical data, including hemodynamic and volumetric measurements (see Table \ref{tbl:data}) \cite{Bonini2025b}. 
Other physiological constraints were enforced by prescribing patient-specific blood volume (via the Nadler equation~\cite{Nadler1962}), scaling coronary perfusion to LV wall mass at 80 mL/min per 100 g~\cite{Barral2011}, and bounding regurgitant orifice sizes according to echocardiographic severity. In previous work \cite{Bonini2025}, we assessed the sensitivity of fitting the 0D model parameters and found that no parameter showed extreme sensitivity, with all changes staying within a reasonable range (below 25\% or within 5 mmHg).
The resulting optimized 0D solutions provided patient-specific cardiovascular dynamics and were used to initialize boundary conditions and circulatory elements in the coupled 3D–0D simulations (Fig. \ref{fig:pipeline}E, Fig. \ref{fig:domains}).

% ---------------------
\subsection{Hemodynamic Modeling} \label{sec:cfd}
To compute patient-specific blood flow and pressure fields, computational fluid dynamics simulations were carried out using the finite-element solver $\boldsymbol{\mathcal{C}}$\textnormal{\textbf{Heart}}~\cite{Lee}. Blood velocity (\textbf{v}) and pressure (\textit{p}) were obtained by solving the Arbitrary Lagrangian–Eulerian Navier–Stokes equations over the time-dependent computational domain $\Omega(t)$~\cite{Balmus2020,Hessenthaler2017}. The governing equations are expressed as:
\begin{align}
    \rho\partial_t \bv 
       + \rho(\bv - \bvg ) \cdot \nabla \bv 
       - \nabla \cdot \boldsymbol{\sigma}
       + \gamma_v \Phi_{v} (\mathbf{v} - \hat{\mathbf{v}})
    & =  \boldsymbol{0}, && \text{on } \Omega (t),    
           \label{eq:sf-ns1} \\
    % -------------------------           
    \nabla\cdot \bv & =  0, && \text{on } \Omega(t),
           \label{eq:sf-ns2} \\ 
    % -------------------------           
    \bv & =  
       \hat{\bv},
       && \text{on } \Gamma_{W}(t), 
           \label{eq:sf-ns3} 
\end{align}
\noindent where, $\rho = $1.06~mm$^3$/g is the fluid density, $\boldsymbol{\sigma} = \mu_f(\nabla\textbf{v} + \nabla\textbf{v}^T)- p\textbf{I}$ is the fluid Cauchy stress tensor, and $\mu_f = 0.004$~Pa$\cdot$s is the bulk viscosity. Prior to running the CFD simulations, a linear-elastic Arbitrary Lagrangian–Eulerian (ALE) problem is solved to solve for the mesh motion ($\hat{\bv}$) \cite{Balmus2020}. 

\begin{figure}[!htb]
    \centering
    \includegraphics[width = 0.7\textwidth]{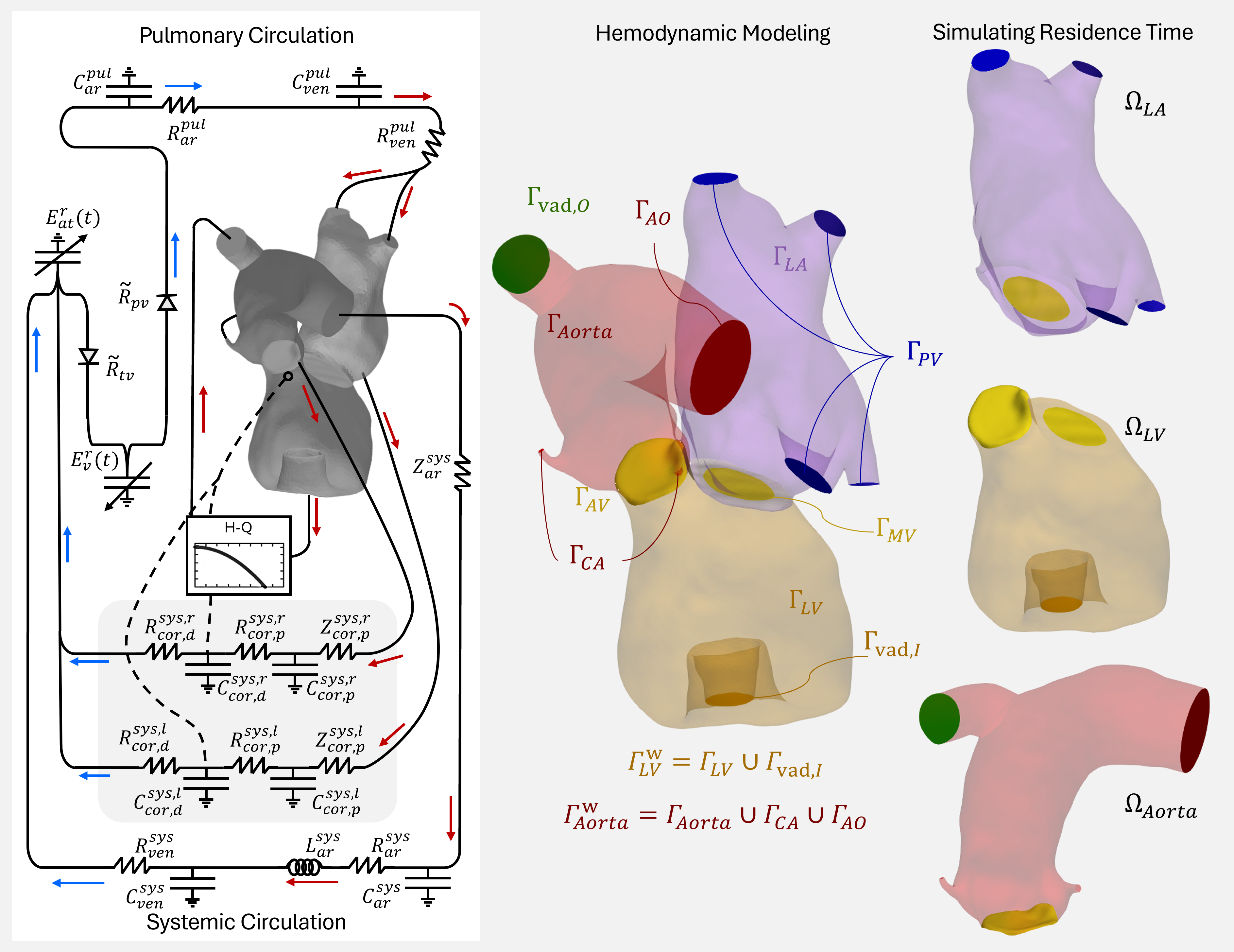}
    \caption{(\textbf{Left}) Illustration of the fully coupled 3D--0D closed-loop cardiovascular model. The LVAD is modeled as a 0D component. (\textbf{Right}) Domains and surfaces for solving the hemodynamic problem and the residence time problem.
    %Comparison of the 3D surface mesh at end-diastole (gray) and end-systole (red), demonstrating the limited deformation characteristic of a failing left heart supported by an LVAD. 
    The 0D coupling interfaces include the aortic root outflow ($\Gamma_{AO}$), pulmonary venous inlets ($\Gamma_{PV}$), and coronary arteries ($\Gamma_{CA}$). Valve regions, shown in yellow, correspond to the mitral ($\Gamma_{MV}$) and aortic ($\Gamma_{AV}$) orifices. The LVAD inflow and outflow cannulae are connected through the 0D LVAD model ($\Gamma_{\text{vad},I}$, $\Gamma_{\text{vad},O}$). The subdomains used for the residence problem are shown in the far right column.
    \textit{Abbreviations:} $R$ or $Z$, resistance; $C$, capacitance; $L$, inductance; $E$, elastance; \textit{pul}, pulmonary; \textit{sys}, systemic; \textit{ven}, veins; \textit{ar}, arteries; \textit{cor}, coronary; \textit{r}, right; \textit{l}, left; \textit{p}, proximal; \textit{d}, distal; \textit{v}, ventricle; \textit{at}, atrium; \textit{tv}, tricuspid valve; \textit{pv}, pulmonary valve; \textit{t}, time; LA, left atrium; LV, left ventricle; \textit{vol}, volume; \textit{AO}, aortic outflow; \textit{PV}, pulmonary veins.} 
    \label{fig:domains}
\end{figure}

To model flow through the mitral and aortic valves, we incorporated a dynamic resistance $(\Phi)$ at the valve surfaces (Eqn. \ref{eq:sf-ns1}). Here, $ \gamma_v $ denotes the surface Dirac delta function defined for the aortic and mitral valve surfaces ($\Gamma_{MV},\Gamma_{AV}$). Each valve plane was assigned a spatially and temporally varying resistance that was determined from the transvalvular pressure gradient and flow direction. Based on the solution, the resistance parameter, $\Phi$, was assigned either a large value ($10^6$ Pa) to impede flow or a value of zero to permit flow across the valve plane. By varying $\Phi$ both spatially on the valve plane and temporally over the cardiac cycle, the model represented the dynamic opening and closing behavior of the mitral and aortic valve orifices. 
% When the pressure gradient favored forward flow, the valve opened, whereas backward flow or negative pressure gradients caused the valve to close.
The mitral valve opening shape was represented as an elliptical orifice based on the patient's two-dimensional transthoracic echocardiogram images, while the aortic valve opening and closing shape was modeled as circular.
For regurgitant valves, nodes corresponding to the patient-specific regurgitant orifice area were maintained at low resistance throughout valve closure, permitting backward flow and reproducing the prescribed degree of valvular insufficiency. To model valve repair, this constraint was removed, allowing the valves to fully close. This method allowed the valve state to emerge dynamically throughout the cardiac cycle rather than being prescribed at fixed times. Additional details are provided in Bonini et al.~\cite{Bonini2025b}.

In the model, the LVAD is represented as a zero-dimensional component (see Fig. \ref{fig:domains}) whose flow depends on a nonlinear relationship between the pressure difference across the device ($\Delta P_{\text{vad}}$) and the flow rate ($Q_{\text{vad}}$), expressed as:
    \begin{align}
        HQ(Q_\text{vad}) := A + B Q_{\text{vad}} + C Q_{\text{vad}}^2 + D Q_{\text{vad}}^3=\Delta P_{\text{vad}}, 
        \label{eq:hqcurve}
    \end{align}
    where the coefficients $A$, $B$, $C$, and $D$ are fitted to the manufacturer-provided H–Q curve~\cite{Santiago2021}.
    
This ensures that the pump delivers flow consistent with the pressure difference across it. Each device has its own set of curves and the patient’s pump speed determines the specific curve used. 
To account for patient-specific pathology, an additional resistance term ($E$) was introduced for Patient E to represent outflow graft obstruction reported in the patient’s medical records. The pressure-flow relationship was therefore modified to:
\begin{align}
    HQ(Q_\text{vad}) := A + B Q_{\text{vad}} + C Q_{\text{vad}}^2 + D Q_{\text{vad}}^3 + EQ_{\text{vad}} = \Delta P_{\text{vad}}.
\end{align}

This additional resistance was estimated as part of the 0D parameter-fitting process, resulting in a total of $n=24$ adjustable parameters. This approach enabled the model to capture the hemodynamic impact of the obstruction.

\subsection{Simulating Residence Time}
Once the ALE Navier-Stokes solution was completed, it was then used to identify regions of high flow recirculation, which are associated with elevated thrombosis risk \cite{Neidlin2021,Esmaily-Moghadam2013}. 
To evaluate this, we calculated the residence time, defined as the duration a fluid particle remains within the 3D model (\ref{fig:pipeline}F). 
This quantity was formulated as a field, allowing spatial mapping of areas where blood may stagnate. 
To determine the residence time, we solved the following advection-diffusion equations \cite{Esmaily-Moghadam2013}:
\begin{align}    % RESIDENCE TIME
    \textbf{Left Heart Simulation}& \label{eq:rtlh}
    \\
    \partial_{t}\phi + \bs{\omega}\cdot\nabla\phi 
        - \nabla\cdot\boldsymbol{D}\nabla\phi
        & = 1,
        && \text{on } \Omega^t,
        \\ 
    \boldsymbol{D}\nabla\phi \cdot \bn & = 0,
        && \text{on } \Gamma^{\text{w},t}_{LV} \cup \Gamma^{t}_{LA} \cup \Gamma^{\text{w},t}_{Aorta},
         \\
     % ------- VAD Conditions --------
    \phi - E(\phi,\Gamma^t_{\text{vad},I})
    & = 0,  
        && \text{on } \Gamma^t_{\text{vad},O}, \label{eq:rt-vad1}
         \\
     % ------- PV Conditions --------
     H(\omega_n) \boldsymbol{D}\nabla\phi\cdot \bn + (1-H(\omega_n)) \phi  & = 0, 
        && \text{on } \Gamma^t_{PV},\label{eq:rt-pv} 
        \\
     % ------- MV Conditions --------
     H(|\omega_n|-\kappa) \boldsymbol{D}\nabla\phi\cdot \bn + (1-H(|\omega_n|-\kappa)) [\![ \phi ]\!] & = 0, 
         && \text{on } \Gamma^t_{MV} \cup \Gamma^t_{AV}, \label{eq:rt-vlv1} \\
% \end{align}
    % \nonumber \\ %\vspace{1cm}
% \begin{align}
    \textbf{Isolated Aorta Simulation} \label{eq:rtao}
    \\
    \partial_{t}\phi + \bs{\omega}\cdot\nabla\phi 
        - \nabla\cdot\boldsymbol{D}\nabla\phi 
        & = 1,
        && \text{on } \Omega^{t}_{\text{Aorta}}, 
         \\ 
    \boldsymbol{D}\nabla\phi\cdot \bn& = 0,
        && \text{on } \Gamma^{\text{w},t}_{Aorta}, 
         \\
    \phi& = 
        0, 
        && \text{on } \Gamma^t_{\text{vad},O},\label{eq:rt-vad2}
        \\
        H(|\omega_n|-\kappa) \boldsymbol{D}\nabla\phi\cdot \bn + (1-H(|\omega_n|-\kappa)) \phi & = 
        0,
        && \text{on } \Gamma^t_{AV}, \label{eq:rt-av}
\end{align}

\noindent where $\bs{\omega}=(\bv-\bvg)$ is the advective velocity field obtained from the Navier-Stokes solution, $\boldsymbol{D}$ is the diffusivity tensor, $\bn$ is the outward normal, $ \omega_n = \bs{\omega} \cdot \bn $ is the normal velocity, $E(a,\Gamma) = |\Gamma|^{-1} \int_\Gamma a dA $ is the mean value operator applied to a field $ a $ over a boundary $ \Gamma $, $H$ is the Heaviside function, $ [\![ \phi ]\!] $ evaluates the jump across valve boundaries, and $\kappa=10.0$~mm/s, which was selected to provide a stable and physically reasonable solution. 
The normal for the MV points into the LV, and the normal for the AV points into the aorta. A constant source term is included so that the residence time field increases steadily over time unless fluid is advected away and replaced by new inflow. 
The advective field from the third cardiac cycle of the fluid simulation is used as the prescribed velocity for this calculation. 
In this study, the physical diffusion coefficient ($\boldsymbol{D}$) is set to zero because it is negligible \cite{Esmaily-Moghadam2013}. 
The residence time equations were solved using a finite element formulation with linear interpolation of $\phi$ over the computational mesh.
To suppress spurious oscillations near sharp gradients, an isotropic Streamline-Orthogonal Laplacian Diffusion (SOLD) stabilization scheme was applied, introducing artificial diffusivity following work by Lynch et al. and Do Carmo and Galeao et al. \cite{lynch2020numerical,do1991feedback,galeao1988consistent}.

Residence time was evaluated in two ways. 
First, it was computed throughout the entire left heart 3D volume \ref{eq:rtlh}. This problem was solved in discrete domains ($\Omega_{LV}^t\cup\Omega_{LA}^t\cup\Omega_{Aorta}^t$), so that the residence time could be independent across valve planes during valve closure (Eq.~\ref{eq:rt-vlv1}). 
The residence time on the LVAD outflow face ($\Gamma_{\text{vad},O}$) was prescribed as the mean residence time of blood entering the LVAD (Eq.~\ref{eq:rt-vad1}).
The residence time field, $\phi$, was set to zero at the pulmonary venous faces ($\Gamma_{\text{PV}}$) on nodes that had inflow into the left atrium (Eq.~\ref{eq:rt-pv}). 
In the second evaluation \ref{eq:rtao}, residence time was calculated only within the aortic volume ($\Omega_{Aorta}$) to isolate the effect of aortic washout and remove the influence of blood transit through the LV and LA, which differs across models. 
In this case, residence time was set to zero for any inflow through the aortic valve face (Eq.~\ref{eq:rt-av}) and the LVAD outflow face (Eq.~\ref{eq:rt-vad2}).
Each residence time simulation was run for 20 cardiac cycles, with less than 5\% change in mean residence time between successive cycles, ensuring temporal convergence of the residence time metrics.

% ---------------------
\subsection{Simulating Left Heart Valve Repair} \label{sec:vlvrepair}
In this study, we simulated repair of the aortic valve, the mitral valve, and both valves simultaneously. 
While valve repair in vivo would elicit patient-specific cardiovascular adaptations, the extent and timescale of these responses are not well characterized for each individual and were not explicitly modeled here.
To focus on the direct hemodynamic consequences of each simulated repair, minimal recalibrations of the 0D model were performed to maintain physiologic behavior.
Specifically, systemic resistances ($R_{ar}^{sys},~Z_{ar}^{sys}$) were adjusted to preserve the pre- to post-repair systolic and diastolic systemic pressures.
This is consistent with changes observed clinically in a subgroup of LVAD patients following transcatheter aortic valve replacement. More information on this can be found in the Supplement, \ref{supp:bp}. 
In a subset of simulated MV\&AV repair cases, without model recalibration, the simulations produced negative pulmonary pressures. We attribute this to excessively high LVAD speeds relative to available preload after simulated valve repair. In vivo, such conditions can lead to ventricular suction or collapse events \cite{Rocchi2025}. However, because our model prescribes ventricular boundary motion, geometric collapse cannot occur. Instead, we observed upstream pressure drops in the pulmonary veins, resulting in non-physiologic negative pulmonary pressures.
To correct this, we modestly reduced LVAD pump speed to maintain non-negative pulmonary pressures and a physiologic cardiac index (2–2.5 L/min/m$^2$), consistent with clinical management guidelines \cite{Imamura2019}. The final calibrated parameters are summarized in Supplementary Table \ref{tab:newparams}.

\subsection{Quantitative Analyses} \label{sec:quant}

To comprehensively assess the hemodynamic consequences of each simulated valve intervention, we extracted a series of clinically relevant metrics from the patient‐specific computational model results (see Table~\ref{tab:quant_metrics}). 
These quantities were selected to quantify right ventricular response, pulmonary vascular load, global circulatory performance, and regions of potential flow stasis or regurgitation. 
Together, they provide a multidimensional view of how simulated valve repair and LVAD operation may influence local and global hemodynamics.  
Specifically, we examined several metrics that capture global and regional cardiac function.
Right ventricular ejection fraction (RVEF) serves as an index of global RV systolic performance.
RV–pulmonary artery coupling ($E_{\text{max}}^{RV}$/Ea) quantifies ventricular–vascular interaction and contractile efficiency, while the pulmonary artery pulsatility index (PAPi) reflects RV preload responsiveness and failure risk.
Mean pulmonary artery pressure (mPAP) serves as a measure of RV afterload and a marker of pulmonary hypertension.
Forward aortic valve flow characterizes aortic valve opening, where in LVAD patients, regular opening reduces commissural fusion and the progression of aortic insufficiency. 
The volumetric flux across valve surfaces was calculated with the following equation: 
\begin{align*}
    Q_{i}=\int_{\Gamma_{i}} (\mathbf{v}-\mathbf{\hat{v}})\cdot \mathbf{n}~dA \quad i={AV,MV}
\end{align*}

\noexpand Then the forward aortic valve flow was calculated by summing all positive $Q_{AV}(t)$ as shown in Table \ref{tab:quant_metrics}.
Finally, relative residence time (rRT) was evaluated, which quantifies intracardiac blood stasis.

\begin{table}[!htb]
\small
\renewcommand{\arraystretch}{1.6} % Increases row height by 1.5x
\centering
\begin{tabular}{ | >{\centering\arraybackslash} m{4.1cm}| >{\centering\arraybackslash}m{4.7cm}|>{\centering\arraybackslash}m{6.2cm}|}
\hline
\textbf{Variable} & \textbf{Equation} & \textbf{Description} \\
\hline
Right Ventricular Ejection Fraction (RVEF) &
    $\displaystyle \frac{\text{RVEDV} - \text{RVESV}}{\text{RVEDV}}$ &
    RVEF$<$45\% defined as RV dysfunction \cite{J.MauricioDelRio2018} 
\\[1ex] \hline
RV-PA Coupling $(E_{\text{max}}^{RV}/Ea)$ &
    $\displaystyle \frac{E_{\text{max}}^{RV}}{(sPAP-dPAP)/RVSV_f} $ &
    Relationship between RV contractility and RV afterload. The Optimal ratio is between 1.5 and 2.0 \cite{He2023,Jone2019}.   
\\[1ex] \hline
Pulmonary Artery Pulsatility Index (PAPi) & 
    \normalsize $\displaystyle \frac{\mathit{sPAP - dPAP}}{\mathit{CVP}}$ &
    Reflects RV preload and afterload; PAPi$<$1.56 suggests risk of RVF \cite{Wei2024}   
\\[1ex] \hline
Mean Pulmonary Artery Pressure (mPAP) &
    $\displaystyle \frac{1}{T}\int_{0}^{T}PAP(t)\,dt$ &
    A mPAP$> = $25mmHg signifies PHT; sPAP$>$50mmHg is severe PHT and is a determinant of RV dysfunction  \cite{LeTourneau2013}. 
\\[1ex] \hline
Forward Aortic Valve Flow &
    $\displaystyle \int_{0}^{T} Q_{\mathrm{AV}}^{+}(t)\,dt$ &
    Net forward volumetric flow through the aortic valve over one cardiac cycle, $T$. 
\\[1ex]\hline
% Mitral Regurgitant Fraction (MRF) &
%     $\displaystyle \frac{\int_{0}^{T} Q_{\mathrm{MV}}^{-}(t)\,dt}{\int_{0}^{T} Q_{\mathrm{MV}}^{+}(t)\,dt}$  &
%     Severity: mild, MRF$<$30\%; moderate, MRF = 30-49\%;  severe, MRF$>$50\% 
% \\[1ex]\hline
Relative Residence Time (rRT) &
    $\phi/T$ &
    Normalized residence time field ($\phi$). Values range from 0 (rapid washout) to 1 (fluid remains for the entire simulation, $T$). 
\\[1ex] \hline
\end{tabular}
\caption{Hemodynamic Variables Quantified. 
Abbreviations: \textit{RVEDV}, right ventricle end diastolic volume; \textit{RVESV}, right ventricle end systolic volume;  \textit{$E_{\text{min}}^{RV}$}, RV end systolic elastance (Table \ref{table:0D values}); \textit{Ea}, pulmonary artery effective arterial elastance; \textit{$RVSV_f$}, RV stroke volume through the pulmonary valve;
\textit{sPAP}, systolic pulmonary artery pressure; \textit{dPAP}, diastolic pulmonary artery pressure; \textit{CVP}, central venous pressure estimated as mean right atrium pressure; \textit{Q$^+$}, positive volumetric flux; \textit{AV}, aortic valve; \textit{MV}, mitral valve.}
\label{tab:quant_metrics}
\end{table}

% ---------------------
% ---------------------
\section{Results} \label{results}
In this work, we modeled 5 LVAD patient-specific cases. 
At their post-LVAD state, all patients had mitral regurgitation and all, except patient E, had aortic insufficiency. 
After modeling patient-specific hemodynamics, we performed \textit{in-silico} mitral valve repair, aortic valve repair, and mitral and aortic valve repair for each patient when appropriate.
The blood velocity and pressure results for Patients A, C, D, and E are in the Supplementary Material (Figures \ref{fig:patA}-\ref{fig:patE}).
The metrics, described in Table \ref{tab:quant_metrics}, were also quantified to evaluate the benefits of simulated left heart valve repair in the patient cohort (see Figures \ref{fig:bars} and \ref{fig:bars_supp}). As patient-specific postoperative validation data were unavailable, these repair scenarios represent \textit{in silico} experimentation rather than validated outcomes.

% ---------------------
\subsection{Patient Specific Results}
Table~\ref{tab:datafit} presents a comparison between model-predicted hemodynamic quantities and corresponding patient measurements, where shaded cells denote the percent deviation from clinical values. Overall, the 3D–0D simulations accurately reproduced key physiological metrics. The mean deviation from patient data was 5.8\% for the 3D–0D model. 

\begin{table}[!htb]
    \centering
    \includegraphics[width=\textwidth]{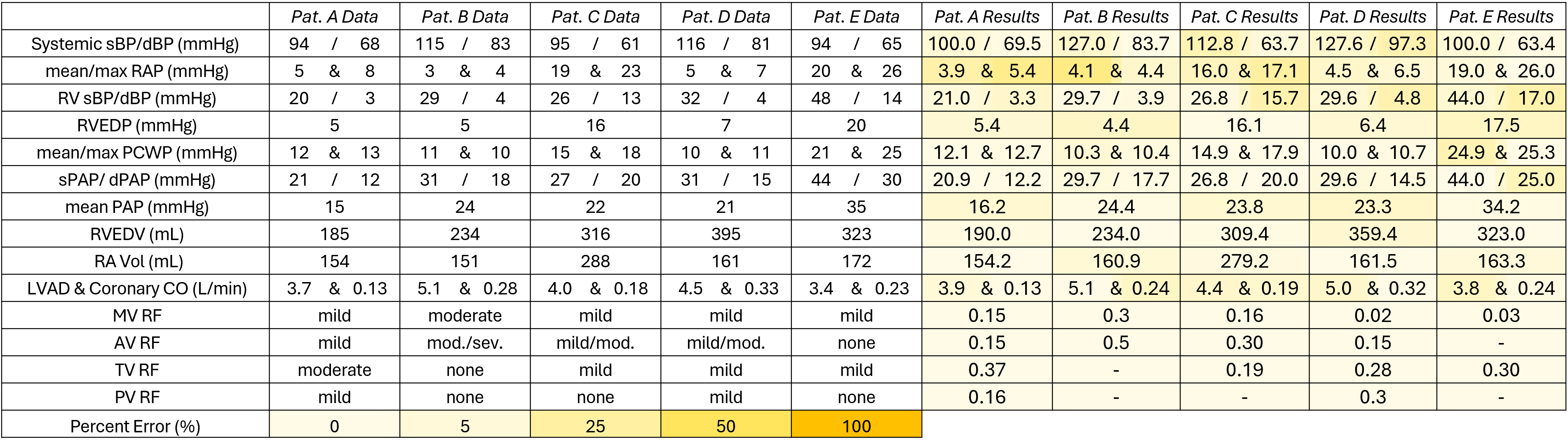}
    \caption{3D-0D Patient Specific Results for Model Validation.  
    The results are shaded to indicate the percent error between the model results 
    and the patient data. The computational model's left atrium pressure was used to 
    Calculate PCWP.  Abbreviations: \textit{CO}, cardiac output; \textit{sBP}, systolic blood pressure; \textit{dBP}, diastolic blood pressure; \textit{RAP}, right atrial pressure; \textit{RVP}, right ventricle pressure; \textit{PAP}, pulmonary artery pressure;\textit{PCWP}, pulmonary capillary wedge pressure; \textit{RVEDP}, right ventricle end diastolic pressure; \textit{RF}, regurgitant fraction; \textit{RVEDV}, right ventricle end diastolic volume; \textit{Vol}, Volume.}  \label{tab:datafit}
\end{table}

\begin{figure}[hbt!]
    \centering
    \includegraphics[width=0.9\linewidth]{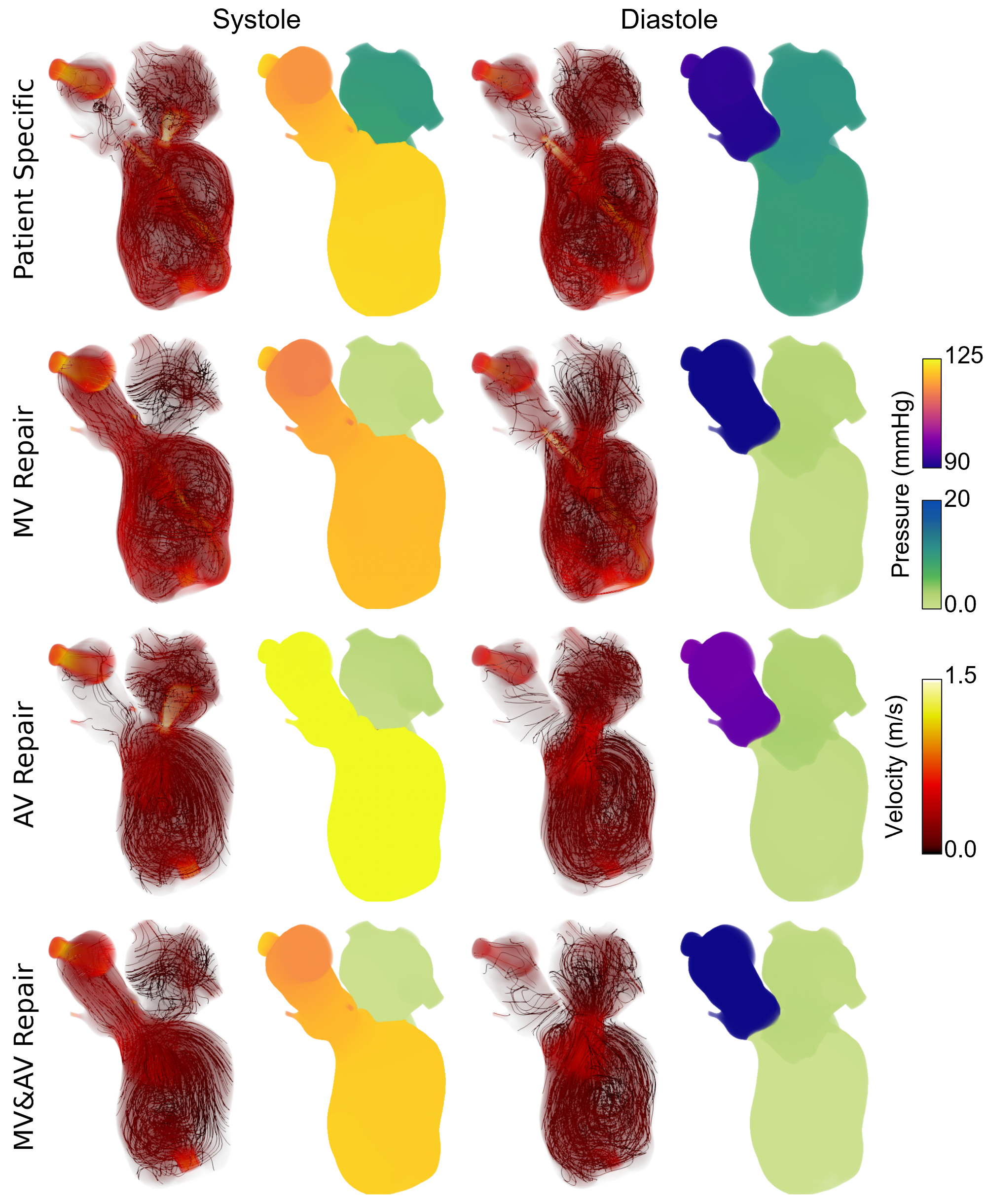}
    \caption{Blood flow and pressure plotted at peak systole and end diastole for Patient B and the results of simulated MV repair, AV repair, and MV\&AV repair. Similar figures for patients A, C, D, and E can be found in the supplementary material (Sec. \ref{sec:supp_res}).}
    \label{fig:patB}
    % \vspace{-1cm}
\end{figure}

Across the cohort (Patients~A–E), the simulations captured distinct hemodynamic characteristics reflecting the varying severity of valve disease.
Patient~B exhibited the most pronounced valvular dysfunction, with moderate aortic and mitral regurgitation, whereas the remaining patients demonstrated only mild regurgitant lesions.
% Furthermore, left ventricular and aortic pressure profiles varied between cases: Patient's B and C showed high LV filling pressures, while Patient A had the lowest LV pressures (Fig. \ref{fig:patA} -- diastole). 
The degree of forward flow through the aortic valve also differed substantially, with Patient~D demonstrating the greatest antegrade ejection, Patients B and E showing moderate flow, Patient~C exhibiting minimal forward flow, and Patient~A showing complete absence of aortic valve opening (Figs. \ref{fig:patB}, \ref{fig:patA}-\ref{fig:patE} -- systole). These trends highlight the hemodynamic heterogeneity within the LVAD population.

% ---------------------
\subsection{simulated Mitral Valve Repair}
\begin{figure}[!htb]
    \centering
    \includegraphics[width = 0.75\textwidth]{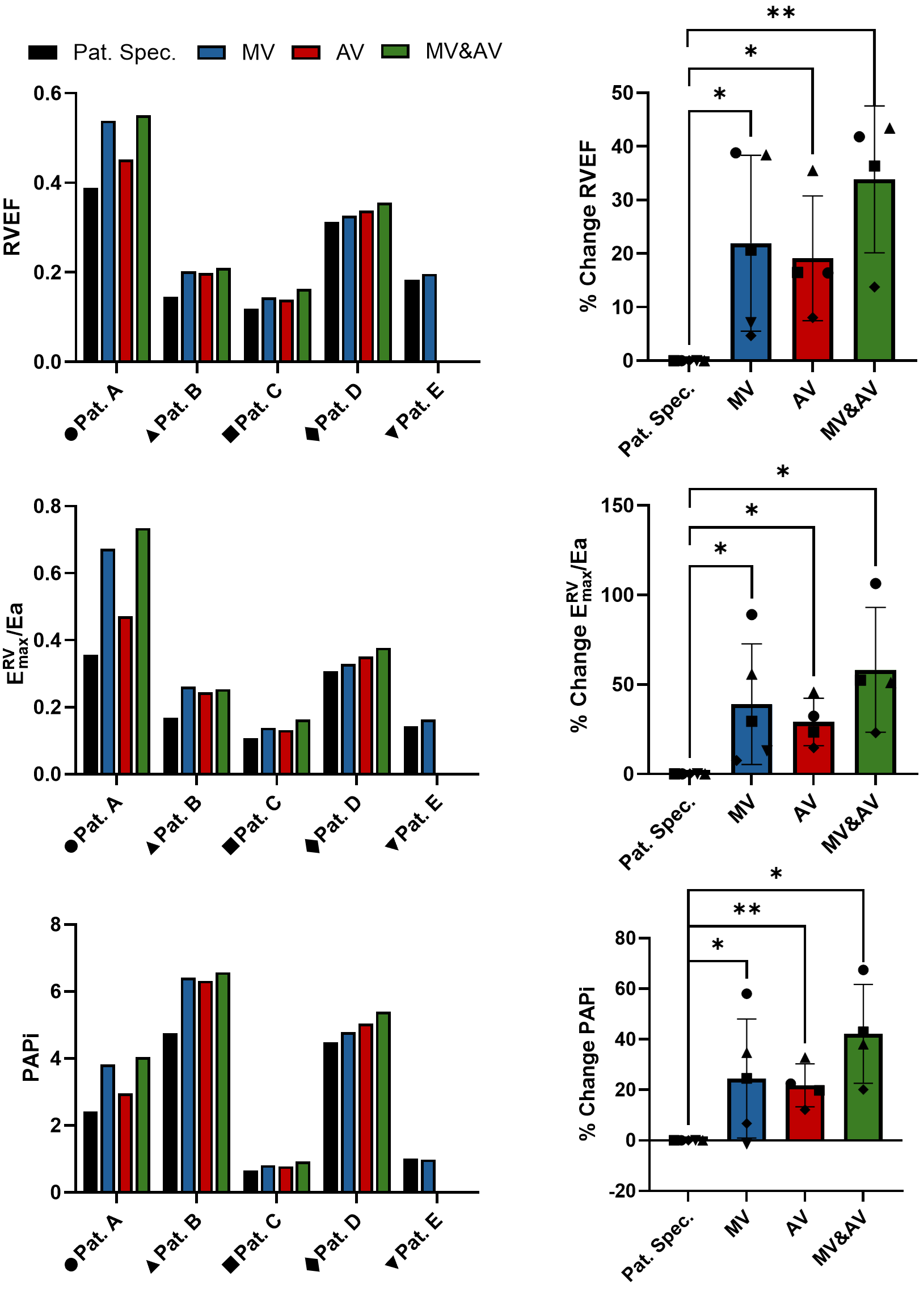}
    \vspace{-.2cm}
    \caption{Right Ventricle Metrics. The values for each metric are plotted on the left, grouped by patient. Black bars are the patient-specific results, blue bars are the results for MV repair, red bars are for AV repair, and green bars are for AV\&MV repair. The plots on the right are showing the average percent increase in each value ($y$) compared to the patient-specific case ($y_{ps}$) for all patients, $(y-y_{ps})/y_{ps} \cdot 100$. And the significance of the average percent increase with respect to the patient-specific cases ($^{**}p<0.01$ and $^{*}p<0.05$). The data is plotted with circles for Patient A, triangles for Patient B, squares for Patient C, diamonds for Patient D, and upside down triangles for Patient E.}
    \label{fig:bars}
\end{figure}

\textit{In silico} mitral valve repair yielded physiologically consistent hemodynamic improvements across all five patient-specific models. Regurgitant flow through the mitral valve was successfully reduced in each case, with retrograde velocity magnitudes in the left atrium substantially diminished relative to pre-repair conditions. On average, the forward cardiac output increased 21.2$\pm 15.91$\% (\textit{p}~=~0.02). The reduction in mitral regurgitation was consistently associated with lower left atrial and pulmonary venous pressures, along with elevated systolic pressures in the left ventricle following simulated repair (Fig. \ref{fig:patB} -- diastole). A one-tailed t-test (n = 5) was performed on the percent change from the patient-specific state to the post–valve-repair condition.
During systole, all models demonstrated left ventricular pressures exceeding those in the aorta, resulting in physiological opening of the aortic valve and forward ejection (Figs. \ref{fig:patB}, \ref{fig:patA}, and \ref{fig:patC} -- systole). 
In fact, the forward flow through the aortic valve increased for all patients (Fig. \ref{fig:bars_supp}). 
\textit{In silico} mitral valve repair also reduced left atrial flow disturbances and pulmonary venous congestion, leading to an 2.5\%-45.1\% reduction in peak RV pressure and an 0.4\%-34.7\% decrease in RV end-diastolic volume (Fig.~\ref{fig:pvloops}, blue). On average, the RVEF and PAPi increased by 21.9$\pm $16.42\% (\textit{p}~=~0.02) and 24.5$\pm 23.55$\% (\textit{p}~=~0.04), respectively. Furthermore, the $E_\mathrm{es}/E_\mathrm{a}$ ratio, a measure of RV-PA coupling, increased by 38.9$\pm 33.70$\% (\textit{p}~=~0.03), reflecting improved right ventricular efficiency following simulated repair. These results are summarized in Fig.~\ref{fig:bars}.

% \clearpage

% ---------------------
\subsection{simulated Aortic Valve Repair}
In all cases, \textit{in silico} AV repair (n=4) removed the high backflow into the LV at the aortic valve. This, in turn, decreased the velocity magnitude in the LV. 
An example of such can be seen in Figure \ref{fig:patB}. 
Here, the velocity magnitude is lower during systole and diastole for the case of AV repair than in the patient-specific state. 
simulated aortic valve repair did not improve aortic valve opening or forward flow in any case; however, it produced a modest reduction in mitral regurgitant volume (21.5\% decrease) compared to the patient-specific state.  
The intervention also lowered left ventricular filling pressures, with mean LV end-diastolic pressure decreasing on average 33.6$\pm 11.68$\% (\textit{p}~=~0.005) following \textit{in silico} repair. 
For example, in Fig. \ref{fig:patB}, the diastolic pressure is lower than the patient-specific results.

\begin{figure}[!htb]
    \centering
    \includegraphics[width=1\textwidth]{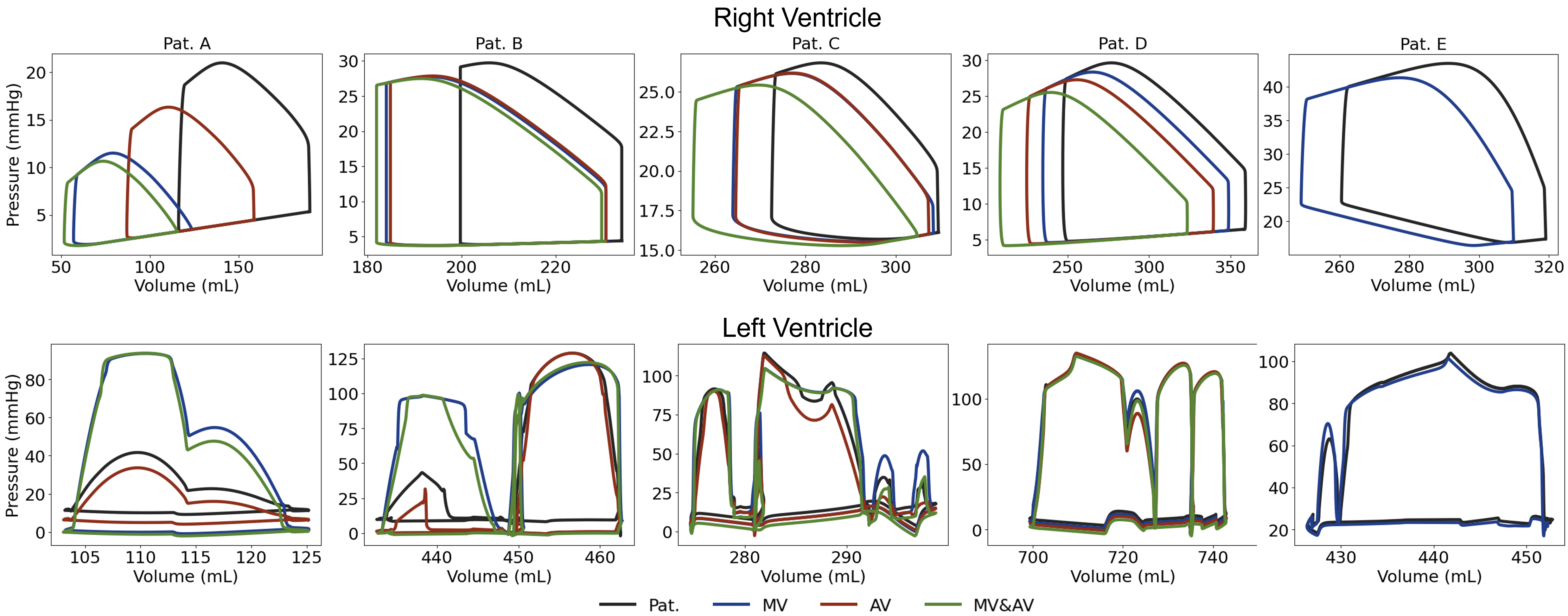} 
    \caption{Right and left ventricle Pressure-Volume Loops for patients A--E. The patient-specific case is plotted in black, the case of MV repair in blue, AV repair in red, and mitral and aortic valve repair in green.}
    \label{fig:pvloops} 
\end{figure}

When evaluating RV metrics, simulated aortic valve repair yielded similar trends to simulated MV repair cases (see Fig. \ref{fig:bars}). 
There was a 2.4-22.2\% and 0.6-16.3\% decrease in peak RV pressure and RV end-diastolic volume, respectively (Fig.~\ref{fig:pvloops}, red). 
Correspondingly, the RVEF and PAPi improved by 19.1$\pm 9.00$\% (\textit{p}~=~0.023) and 21.8$\pm 5.60$\% (\textit{p}~=~0.007), respectively. 
The $E_\mathrm{es}/E_\mathrm{a}$ ratio, also increased by 29.0$\pm 9.17$\% (\textit{p}~=~0.011), signifying enhanced right ventricular performance following \textit{in silico} valve repair. 
Patient B showed greater improvement over the other patients after AV repair, as Patient B had moderate AI and the other patients had mild AI.  

% ---------------------
\subsection{simulated Mitral and Aortic Valve Repair}
When both valves were repaired simultaneously, the hemodynamic improvements were more pronounced. 
Averaged across four patients (A--D), forward cardiac output increased 33.6$\pm 9.72$\% (\textit{p}~=~0.011). Overall, \textit{in silico} valve repair decreased flow velocities within the left heart and aorta. Consistent with the simulated mitral valve repair results, aortic valve opening was restored. Combined \textit{in silico} mitral and aortic valve repair lowered the LV diastolic pressure more than isolated MV or AV \textit{in silico} repair.

Combined mitral and aortic valve \textit{in silico} repair also reduced pulmonary pressures, which may lead to improved RV function. This is seen in the RV P-V loops showing the lowest pressures and volumes for the case of MV and AV repair (Fig.~\ref{fig:pvloops}, green). The average RVEF, PAPi, and $E_\mathrm{es}/E_\mathrm{a}$ ratio increased by 33.8$\pm 3.02$\% (\textit{p}~=~0.008), 42.2$\pm 12.89$\% (\textit{p}~=~0.011), and 58.2$\pm 25.81$\% (\textit{p}~=~0.022) respectively. These findings are summarized in Figure~\ref{fig:bars}.

% ---------------------
\subsection{Residence Time}
\begin{figure}[!htb]
    \centering
    \includegraphics[width=0.75\textwidth]{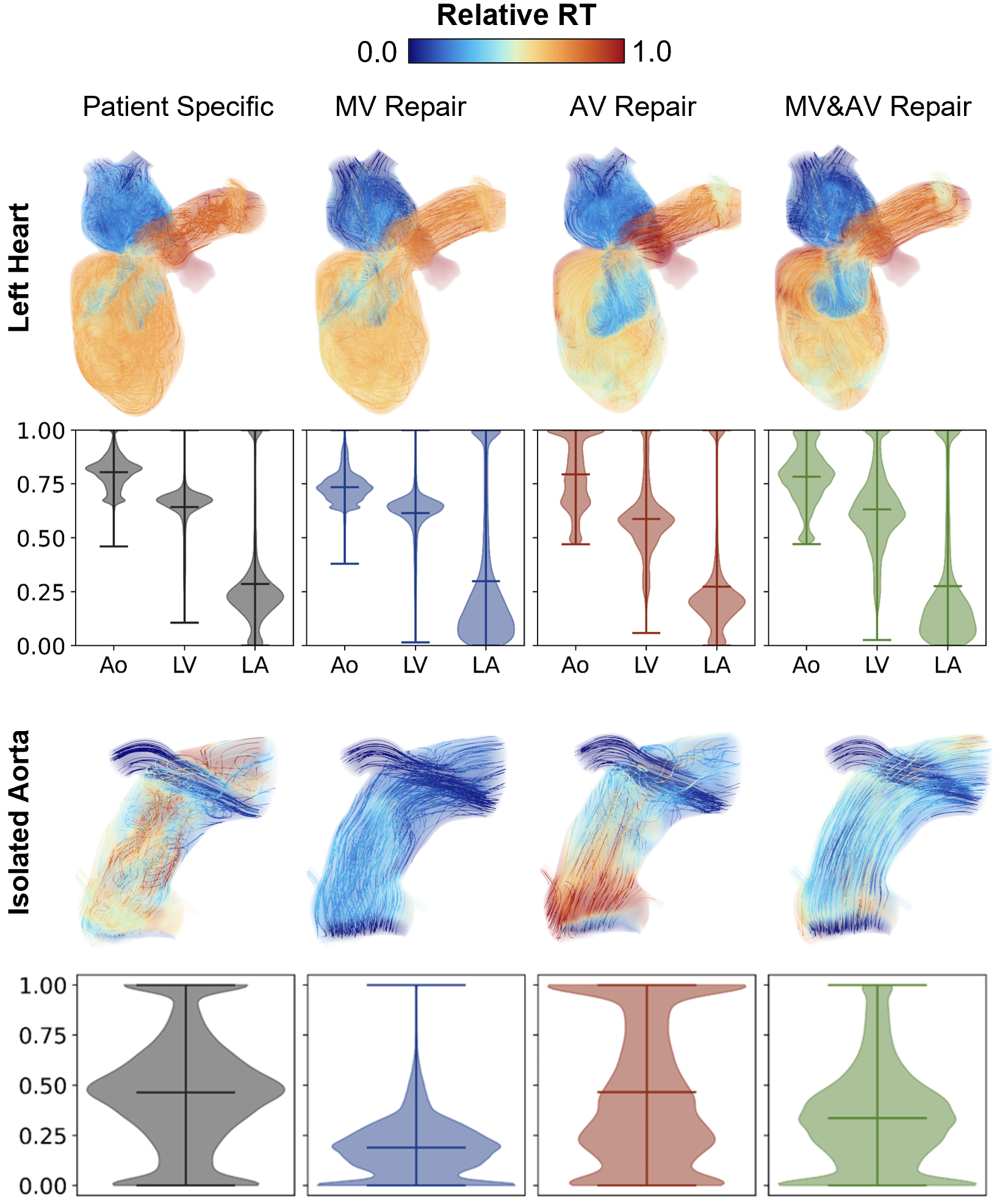} 
    \caption{Relative residence time (rRT) results for Patient~B at peak systole. Row~1 shows the rRT field from the left heart simulation, and Row~2 presents violin plots of the rRT distributions for the aorta ($\Omega_{Aorta}$), left ventricle ($\Omega_{LV}$), and left atrium ($\Omega_{LA}$). Rows~3 and~4 show results from the isolated aorta simulations, including the 3D rRT field (Row~3) and the corresponding violin plot distributions (Row~4). Relative residence time is visualized using volume rendering, with streamlines indicating flow direction and colored according to local rRT values.}
    \label{fig:rtB} 
\end{figure}

The relative residence time (rRT) for Patient~B is shown in Figure~\ref{fig:rtB}. The first row illustrates the spatial distribution of rRT across the left heart and aorta, while the second row shows violin plots of the rRT distributions within each subdomain. Overall, the aorta, particularly the aortic sinuses, exhibited the highest rRT values.
Following simulated mitral valve repair, aortic valve opening improved washout within the aorta. This effect was most pronounced in the isolated aorta case, which showed a 59.3\% reduction in mean rRT. In contrast, simulated AV repair increased the mean aortic rRT slightly (0.34\%) for the isolated aorta case. In the left heart simulation, MV repair and AV repair reduced the LV rRT by 4.4\% and 8.5\%, respectively.
Results for all patients are provided in Supplementary Material~\ref{supp:rt}.

% \clearpage

% ---------------------
% ---------------------
\section{Discussion} \label{discussion}
This study applies a patient-specific computational modeling framework to investigate the hemodynamic consequences of \textit{in silico} mitral and aortic valve repair in LVAD-supported patients. 
By integrating high-resolution three-dimensional flow simulations with a closed-loop lumped-parameter circulation model, this approach captures both local and systemic effects of \textit{in silico} valve intervention under patient-specific conditions. 
In addition to reproducing global hemodynamics, the framework enables spatially resolved quantification of blood residence time (RT) throughout the left heart and aorta, allowing detailed assessment of flow stasis and washout that are not measurable in vivo. 
This combined 3D–0D formulation therefore provides a comprehensive view of how valve competence influences left ventricular and aortic flow, systemic circulation, and the right heart. 
Such an integrated approach is critical in LVAD physiology, where heart failure is rarely confined to the left ventricle. 
Instead, valve dysfunction, altered loading conditions, and ventricular interdependence create tightly coupled interactions between the left and right heart. 
Understanding these relationships through mechanistic modeling provides a unique opportunity to quantify the benefits of valve repair and assess changes in blood stasis.  These insights may help inform LVAD management strategies that are otherwise difficult to evaluate in heterogeneous clinical populations. While postoperative valve repair data were not available for validation, the patient-specific models reproduced clinical data with good agreement. 
This supports the use of these simulations as a physiologically grounded framework for exploratory \textit{in silico} investigation.

% - ----- LV -----
\subsection{Left Heart and Aortic Hemodynamics}
Across all patients, simulated mitral and aortic valve repair markedly improved left heart function, with the most substantial benefits observed following combined MV\&AV repair. Despite inter-patient differences in patient-specific ventricular performance, forward cardiac output (CO) increased following simulated repair for all patients (Fig. \ref{fig:bars_supp}). 
Restoration of forward flow allowed LVAD pump speeds to be reduced in some patients to maintain physiological pulmonary venous pressures and cardiac index. 
This is clinically meaningful, as operating LVADs at lower speeds reduces suction events, pump-induced shear exposure, and may reduce the incidence of RV dysfunction \cite{Couperus2017-lq,Hayward2011-yu,Keenan2022-gh}.
Across all patients, mean left atrial pressure decreased (1.3--10.9 mmHg), consistent with prior findings by Raghunathan et al., who reported similar reductions following transcatheter mitral valve repair in LVAD patients \cite{Raghunathan2021-dx}.

One of the most notable findings was the restoration of aortic valve opening following simulated mitral valve repair. In the patient-specific LVAD-supported state, persistent aortic valve closure was noted in Patient A. 
Even though the increase in forward flow did not reach statistical significance (\textit{p}~=0.16–0.20 across repair types), all patients demonstrated improved aortic ejection (Section \ref{sec:supp_res}). This shows that reducing MR increases LV systolic pressure, potentially enabling transvalvular pressure gradients large enough to open the AV. 
Aortic valve opening is critical as it prevents commissural fusion, and patients with intermittent valve opening have been shown to exhibit a lower incidence of aortic insufficiency \cite{Purohit2018,Noly2020}.
% In contrast, isolated AV repair reduced regurgitant backflow into the LV but did not meaningfully restore AV opening or increase forward flow. 
In this cohort of five patients, isolated AV repair reduced regurgitant backflow into the LV but did not result in an increase in effective AV opening or forward flow. The impact of simulated AV repair on forward flow appeared limited relative to simulated MV intervention, which more directly influenced transmitral filling and downstream washout. 
This highlights that eliminating AI alone does not guarantee systolic aortic valve function. Sufficient LV systolic pressure is required to open the aortic valve, which, as these results show, is primarily achieved through correction of MR.
Interestingly, this relationship between MR reduction and AV opening in LVAD patients has not been extensively investigated clinically. 
Our findings suggest that persistent absence of MR may contribute to more consistent AV opening and potentially protect against AV fusion and AI progression. This warrants further investigation in larger clinical cohorts to evaluate whether correction of MR is protective against AI development.
With combined \textit{in silico} MV\&AV repair, forward CO increased the most (33.6\% on average), and AV opening was consistently restored. This emphasizes the synergistic role of both valves in governing LV pressure generation, aortic outflow, and pump–ventricle interaction.

\subsection{Residence Time}
Differences in RT among patients primarily reflected variation in left atrial and ventricular volumes, consistent with Little’s Law ($\phi_{mean}=V/Q$) \cite{Esmaily-Moghadam2013}, where larger chamber volumes prolong blood residence. 
Patient's~B and D had the greatest LV volumes and exhibited the slowest washout, whereas Patient~A has the smallest LV volume and the quickest washout.
Across all simulations, the regions of highest RT were in the aortic sinuses, which is consistent with previously reported zones of stasis \cite{Wong2014IntraventricularHeart,May-Newman2013-ka}. %—namely, the LV apex near the LVAD inflow cannula, the LV outflow tract, and the aortic sinuses 
simulated valve repair altered the distribution and magnitude of these stagnation regions in distinct ways. 
Mitral valve repair restored aortic valve opening and generated new forward flow paths that refreshed the aortic root, thereby reducing sinus rRT in most cases. This finding agrees with other works that found that persistent AV closure is a known risk factor for aortic root thrombosis \cite{Noly2020,Fried2018-ai}. Thus, improving aortic valve motion through MV repair could help mitigate blood stasis by promoting regular opening and ejection across the aortic valve plane.
In contrast, simulated aortic valve repair showed mixed changes to the mean rRT in the aorta.
In patient A, it increased the RT (23.0\%) within the aortic root despite improving overall hemodynamic efficiency. 
Under patient-specific conditions, AI generated retrograde recirculation from the LVAD outflow graft into the aortic root, which, while hemodynamically unfavorable, enhanced local mixing and paradoxically reduced RT. This aligns with other studies that have found that AI has not been shown to correlate with thromboembolic events \cite{Noly2020}. 
When AI was eliminated, this mixing ceased, resulting in slower sinus washout and slightly elevated RT values. In patients B--D, however, there was a slight decrease in rRT (1.2\%, 5.4\%, 6.7\%).
simulated AV repair resulted in a slightly lower mean rRT within the LV, as the absence of retrograde flow from the aorta eliminated inflow of high-rRT fluid into the ventricle. However, without the high-velocity retrograde jet that previously promoted mixing, localized regions in the LV developed elevated rRT, particularly near the basal segments and left ventricular outflow tract (LVOT), areas previously reported to exhibit high stasis \cite{Wong2014IntraventricularHeart,May-Newman2013-ka,Neidlin2021}.
It is important to emphasize, however, that a lower RT under AI does not imply a favorable condition: AI represents a closed-loop circulation where blood is repeatedly shunted between the aorta and LV, lowering systemic output and pump efficiency. Therefore, while AI may artificially shorten residence time in the aortic root, it does so at the expense of overall cardiac performance. This is reflected in clinical data showing that AI is associated with increased risk of hospitalization and reduced survival \cite{Truby2018AorticRegistry}.
Collectively, these findings show that RT patterns in LVAD patients are driven by valve competence, flow directionality, and LVAD support, and that MV repair may improve hemodynamics and reduce blood stasis by restoring forward aortic flow.

\subsection{Right Ventricle Metrics}
Across the cohort, in silico valve repair generally improved indices of right ventricular performance, underscoring the strong coupling between left‐sided valve competence and RV loading in LVAD support. RVEF is a well-established marker of global RV systolic function, where values below 0.45 indicate dysfunction. It increased in all simulated repair scenarios compared to the patient-specific state. Combined mitral and aortic valve \textit{in silico} repair produced the largest relative improvement, followed by isolated MV, then isolated AV repair. This is likely because it simultaneously restores forward flow and reduces regurgitant volumes across both valves, thereby maintaining low diastolic pressures and restoring pump efficiency. In contrast, isolated MV repair more directly influenced RV metrics compared to AV repair, by normalizing left atrial pressure–volume dynamics, improving RV preload, and lowering pulmonary pressures. Despite inter-patient variability, mean pulmonary artery pressure (mPAP) decreased in all cases, indicating overall unloading of the pulmonary circulation; however, this reduction did not reach statistical significance (\textit{p}~$>$~0.01)(Supplementary~Fig.~\ref{fig:bars_supp}).

The pulmonary artery pulsatility index (PAPi) also increased after simulated valve repair. This metric reflects the RV’s ability to translate preload into forward flow~\cite{Essandoh2022}. This aligns with clinical studies that also saw that PAPi worsened with AI \cite{Bhagra2016-qz}. Prior studies have demonstrated that a low PAPi ($<$1–2.17) is correlated with RV failure both before and after LVAD implantation~\cite{Essandoh2022,Wei2024}. This suggests that even modest increases in PAPi may translate into clinically meaningful improvement. In our simulations, PAPi increased across all simulated repairs, with MV repair producing a higher increase than AV repair, and the greatest enhancement was observed for combined MV\&AV repair. This pattern suggests that mitral competence contributes more directly to RV pulsatility.
Further insight into RV–pulmonary artery (PA) interaction was gained through the ratio of end‐systolic to arterial elastance ($E_{\text{max}}^{RV}$/Ea), a load‐independent index of RV–PA coupling. Optimal coupling typically occurs for $E_{\text{max}}^{RV}$/Ea values between~1.5 and~2.0, whereas uncoupling reflects the inability of the RV to augment contractility in response to increased afterload~\cite{He2023,Stpr2024,Adly2023}. Although all model results fell below this range, all simulated repair scenarios increased $E_{\text{max}}^{RV}$/Ea, showing improved RV–PA coupling. The combined MV\&AV repair achieved the largest average improvement. 

Patient A exhibited a greater improvement in pulmonary hemodynamics compared to the models, which may be attributed to differences in patient-specific pulmonary vascular resistance (PVR). Patient A had a markedly lower PVR (0.07 mmHg$\cdot$s/mL), well below the clinical threshold for pulmonary hypertension (PVR $>$ 0.18 mmHg$\cdot$s/mL) \cite{Thenappan2016-ag}. In contrast, Patients B, C, D, and E demonstrated elevated PVR values of 0.276, 0.166, 0.170, 0.132 mmHg$\cdot$s/mL, respectively. In patients with higher PVR, the pulmonary vasculature may be less capable of accommodating changes in flow following ventricular unloading, resulting in a dampened hemodynamic response. This could explain why these patients experienced smaller reductions in pulmonary pressures and less overall benefit compared to Patient A. 
Furthermore, Patients D and E had very mild MR ($<$2~mL/beat), which further explains why \textit{in silico} repair had a limited impact in these cases. Collectively, these findings highlight the substantial heterogeneity across patients, which can lead to markedly different hemodynamic responses following intervention. This variability may help explain the conflicting conclusions reported in the literature regarding the role of MR, where its impact has been described as both detrimental and, in some contexts, compensatory \cite{Kherallah2024,Kanwar2020,Rad2023,Noly2022}. Taken together, these results suggest that the benefit of valve repair is highly patient-specific, with significant improvements observed in some individuals, while others may experience minimal hemodynamic gain. Future work is needed to evaluate which patients will benefit most from these procedures.

Overall, these findings highlight the interdependence between left‐sided valve function and right heart performance in LVAD‐supported patients. Restoring mitral competence primarily alleviates pulmonary congestion and RV preload stress, while aortic repair further stabilizes ventricular loading conditions and systemic outflow. Collectively, these results suggest that correcting regurgitant lesions in LVAD patients may prevent progressive RV–PA uncoupling and improve RV loading conditions, a finding with potential implications for perioperative valve management and long‐term device outcomes.

% ---------------------
\subsection{Limitations}
Several limitations should be acknowledged when interpreting the findings of this study.
First, the computational framework does not capture dynamic or long-term cardiovascular adaptation. Cardiac motion in the model is prescribed directly from patient imaging, which reproduces subject-specific geometry and kinematics but constrains deformation to those imaged states. As a result, the myocardium cannot adjust its motion in response to altered loading conditions, simulated valve repair, or LVAD speed changes. 
This limitation is partly mitigated in the LVAD setting, where continuous unloading reduces native ventricular stroke volume and contractile excursion, so the magnitude of beat-to-beat myocardial deformation is relatively small \cite{Thohan2005-yr}. 
In fact, for this reason, some CFD models use a rigid domain \cite{Neidlin2021}.
However, our approach still precludes any active myocardial adaptation or structural remodeling that may emerge, particularly over longer timescales.
Although systemic vascular resistance and pump speed were recalibrated after valve repair to avoid unphysiological pressures, the model does not account for long-term global cardiovascular adaptation. In vivo, compensatory processes such as baroreflex control (which is muted in continuous-flow LVAD patients \cite{Purohit2018}), ventricular remodeling, pulmonary vascular adaptation, and right-sided recovery can influence long-term hemodynamics after valve intervention or LVAD speed optimization. However, the difficulty in modeling these multi-scale, adaptive pathways is not unique to this work and remains a challenge for patient-specific cardiovascular simulations. 

Second, valve function was modeled using simplified planar surfaces with prescribed effective orifice areas rather than fully resolved leaflet anatomy. 
This formulation allows physiologic variation in transvalvular resistance throughout the cardiac cycle but does not capture leaflet-level mechanics such as coaptation, tethering, or annular dynamics. Furthermore, the shape of the valve opening and closure area is explicitly defined, which may contribute to the high regurgitant velocities observed in the left atrium and the resulting backflow into the pulmonary veins—features typically associated with severe MR. In addition, the model does not account for perioperative factors that may influence actual valve repair outcomes, such as surgical technique, annuloplasty ring sizing, residual leaflet tethering, or acute changes in ventricular loading following surgery. For these reasons, restoration of aortic valve opening following mitral valve repair should be interpreted as a hemodynamic model hypothesis based on the simulated pressure gradients and forward flow, rather than direct proof of postoperative valve behavior. Despite these simplifications, the model reproduced clinically relevant features of regurgitant behavior, including regurgitant volume, regurgitant fraction, and valve closure timing (within 5\% of the cardiac cycle) \citep{Yoganathan2004}.

The right ventricle is modeled using a reduced-order, time-varying elastance formulation, without explicitly accounting for septal interaction or pericardial effects. Although this approach is widely used and has been validated in prior studies \cite{Comunale2021}, interpretations of right-sided behavior are restricted to overall hemodynamic loading conditions rather than detailed ventricular mechanics and changes in RV function.

In addition, the patient cohort did not include cases of severe MR or severe AI post-LVAD. All modeled patients exhibited only mild or moderate regurgitation following ventricular unloading, and it is this post-LVAD state that the models reflect. 
Future studies with larger cohorts that include patients with persistent severe MR following LVAD implantation will be needed, as the current models may not capture the structural remodeling or maladaptive compensation present in this population. Accordingly, the present study should primarily be interpreted as an investigation into how valve regurgitation influences LVAD hemodynamics, rather than as an assessment of which patients should undergo valve repair.
The patient-specific nature of simulated repair outcomes is evident across the cohort: Patients D and E exhibited only very mild MR and AI ($<$2 mL/beat), resulting in minimal hemodynamic changes following simulated valve repair, while Patient B demonstrated more substantial improvements with repair of moderate AI and MR. These \textit{in silico} predictions suggest that the hemodynamic benefits of valve repair observed here could be even more pronounced in patients with more severe AI or MR, though future validation with larger cohorts will be needed to confirm this. This further underscores the patient-specific nature of repair outcomes and may contribute to variability reported across studies regarding the benefits of valve intervention \cite{Kherallah2024,Kanwar2020,Rad2023,Noly2022}.
Further, the small sample size ($N=5$) limits the ability to draw definitive statistical conclusions. Accordingly, the significance testing performed should be interpreted as exploratory rather than confirmatory. The analysis is intended to identify trends and generate hypotheses, and larger cohorts will be required to establish statistical significance and generalizability. In addition, postoperative multimodal datasets following valve repair in patients with LVAD were not available for this cohort. As a result, the valve repair simulations could not be directly validated against patient-specific data and should therefore be interpreted as exploratory \textit{in silico} experiments. Future work will focus on validating these computational results against postoperative clinical data as such datasets become available.

Finally, there are inherent limitations in the clinical data used for model personalization. The patient datasets were not acquired simultaneously across imaging, hemodynamic, and device measurements, leading to minor temporal misalignment between modalities. Consequently, some values—such as pressures, volumes, and LVAD operating conditions—may not represent the exact same physiological state. Moreover, measurement uncertainty exists in the clinical data itself; for example, LVAD flow estimates can vary by approximately 1 L/min depending on device calibration and physiologic state~\cite{Rocchi2025,Abart2025}. For Patients B and C, dynamic CT imaging was performed at lower LVAD pump speeds to allow aortic valve opening during evaluation for potential transcatheter aortic valve replacement. As a result, the imaging data reflect slightly different loading conditions than the hemodynamic measurements acquired at each patient’s usual operating speed. These factors introduce a degree of variability when comparing model predictions to clinical measurements, but are unavoidable given the multimodal and retrospective nature of the dataset. %.

% ---------------------
% ---------------------
\section{Conclusion} \label{conclusion}
This study presents a patient-specific, image-based computational framework to evaluate the hemodynamic impact of \textit{in silico} mitral and aortic valve repair in LVAD-supported patients. By integrating 3D cardiovascular flow simulations with personalized 0D circulatory models, we were able to isolate the effects of each repair strategy under controlled physiological conditions that are not easily attainable in clinical settings.
Across the cohort, simulated mitral and aortic valve repair improved forward cardiac output and reduced the need for high LVAD pump speeds. Simulated mitral valve repair consistently lowered left atrial and pulmonary venous pressures and enabled aortic valve opening in all patients. While simulated isolated aortic or mitral valve repair reduced retrograde flow and pulmonary pressures, the greatest improvements were achieved when both valves were repaired, reflecting the interdependence of mitral inflow, aortic outflow, and LVAD function.
Residence time analysis further revealed that simulated valve repair improved blood washout in the left heart and aortic root, particularly following mitral valve correction, which restored forward ejection through the aortic valve. These changes reduced stagnation in regions typically associated with thrombosis risk, demonstrating that restoring valve competence could benefit both hemodynamic efficiency and local flow patterns.
These left-sided improvements translated to meaningful changes in right heart loading, including reductions in RV pressures, enhanced RV–pulmonary artery coupling, and increased PAPi. These findings highlight that correcting mitral and aortic insufficiency could not only improves left heart hemodynamics and washout efficiency but may also mitigate progressive right ventricular dysfunction in the LVAD population.

\section*{Credit authorship contribution statement}
% ADD
\textbf{Mia Bonini}: Writing – original draft, Visualization, Software, Methodology, Investigation, Conceptualization, Writing – review \& editing.
\textbf{Michael Ferguson:} Investigation.
\textbf{Marc Hirschvogel:} Software, Methodology, Writing – review \& editing. 
\textbf{Maximilian Balmus:} Software.
\textbf{Paul C. Tang:} Data curation, Writing – review \& editing, Conceptualization.
\textbf{Francis Pagani:} Data curation, Writing – review \& editing, Conceptualization.
\textbf{David A. Nordsletten}: Writing – review \& editing, Supervision, Software, Resources, Methodology, Conceptualization.

\section*{Declaration of competing interest}
Dr. Pagani is a non-compensated ad-hoc scientific advisor for Abbott, BrioHealth Solutions,  and FineHeart. Dr. Pagani is a non-compensated medical monitor for Abiomed and receives grant funding from the National Heart, Lung, and Blood Institute and the Agency for Healthcare Research and Quality. Dr. Pagani receives partial salary support from Blue Cross / Blue Shield of Michigan as Associate Director of the Michigan Society of Thoracic and Cardiovascular Surgeons Quality Collaborative. The other authors declare that they have no known competing financial interests or personal relationships that could have appeared to influence the work reported in this paper.

\section*{Ethics}
All data collection and analysis were conducted under IRB-approved protocol HUM00196629 (approved April 2021)

\section*{Acknowledgments}
This work was supported by the National Institute of Health National Heart, Lung, \& Blood Institute [No. R01HL170059]. This research was supported in part through computational resources and services provided by Advanced Research Computing at the University of Michigan, Ann Arbor.

\clearpage

\renewcommand{\thesection}{SM\arabic{section}}
\setcounter{section}{0}%
\setcounter{subsection}{0}%
\setcounter{equation}{0}
\setcounter{figure}{0}
\setcounter{table}{0}
\renewcommand{\theequation}{SM\arabic{equation}}
\renewcommand{\thefigure}{SM\arabic{figure}}
\renewcommand{\thetable}{SM\arabic{table}}

\section{Supplementary Material}\label{supp}

\
\subsection{Mesh Independence Study} \label{supp:mesh}
\begin{figure}[htb!]
    \centering
    \includegraphics[width=0.7\linewidth]{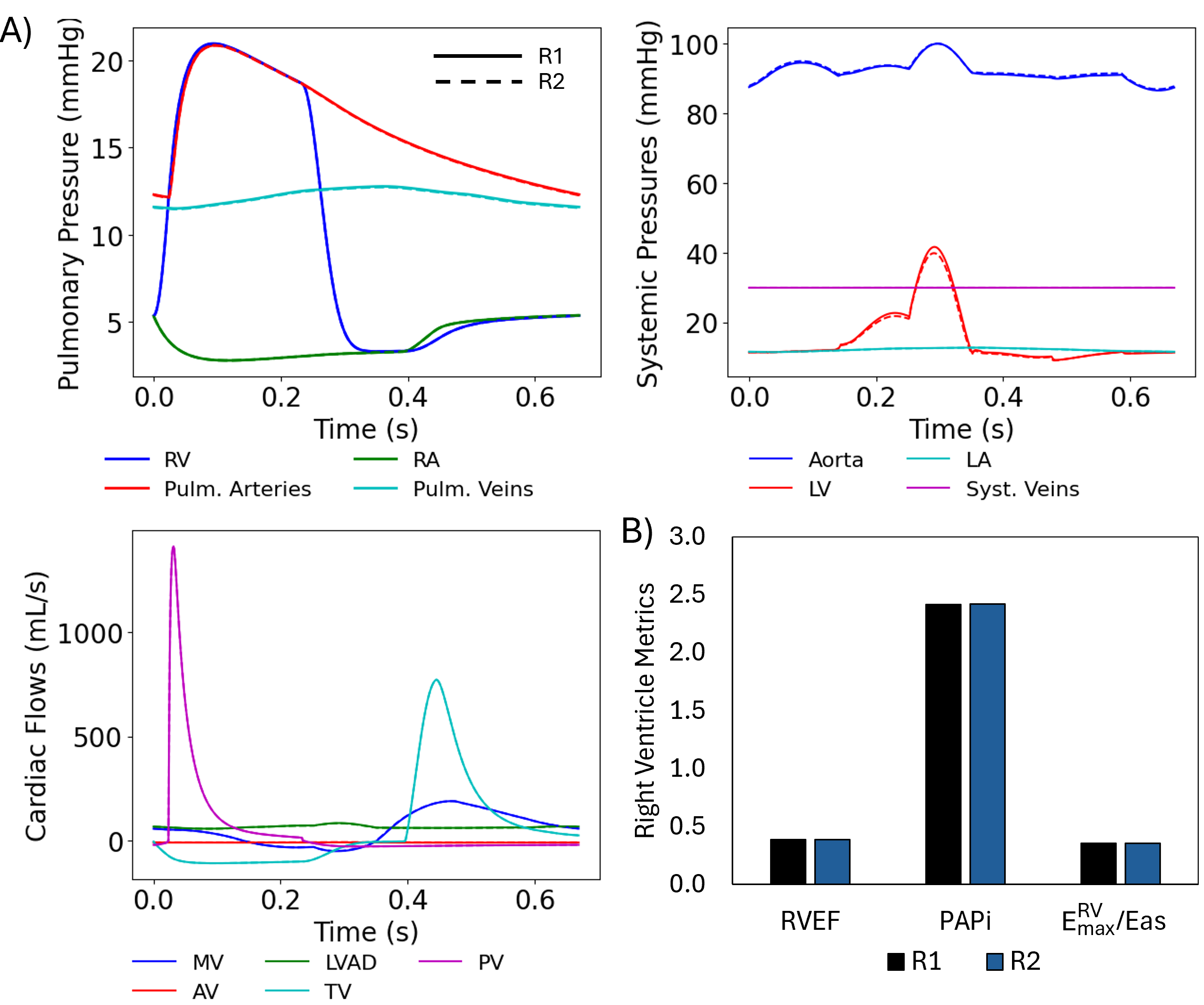}
    \caption{A) Comparison of pressure and volumetric flows between mesh R1 and R2. R1 is plotted with solid lines and R2 is plotted with dashed lines. B) Comparison of RV metrics between R1 (black) and R2 (blue) meshes.}
    \label{fig:mesh2}
\end{figure}

To examine the convergence of our mesh on the metrics being evaluated, we examined two mesh sizes for Patient A (see Table \ref{tab:meshsize}). 
\begin{table}[htb!]
    \renewcommand{\arraystretch}{1.3}
    \centering
    \begin{tabular}{
    >{\centering\arraybackslash}m{4cm}
    >{\centering\arraybackslash}m{3cm}
    >{\centering\arraybackslash}m{3cm}
    >{\centering\arraybackslash}m{3cm}}
    \hline
    \textbf{Mesh} & \textit{$N_x$} & \textit{$N_e$} & \textit{h} [mm] \\ \hline
    \textbf{Course Mesh, R1} & 394,307   & 2,237,460 & 1.0 \\ 
    \textbf{Fine Mesh, R2}   & 1,679,220 & 9,795,569 & 0.5 \\ \hline
    \end{tabular}
    \caption{Number of nodes ($N_x$), number of elements ($N_e$), and average mesh edge length (\textit{h}) for mesh refinement study.}
    \label{tab:meshsize}
\end{table}
We compared the pressure and volumetric flow in the 3D-0D model (average difference of 2.74\%) showing very similar results between the two mesh sizes (see Fig. \ref{fig:mesh2}A). We also compared the RV variables quantified in Table \ref{tab:quant_metrics} which is summarized in Figure \ref{fig:mesh2}B. The percent differences range from 0.12\% to 0.23\% suggesting minimal impact on the quantification of these measures as the mesh size changes significantly.

\begin{figure}[hbt!]
    \centering
    \includegraphics[width=1.0\linewidth]{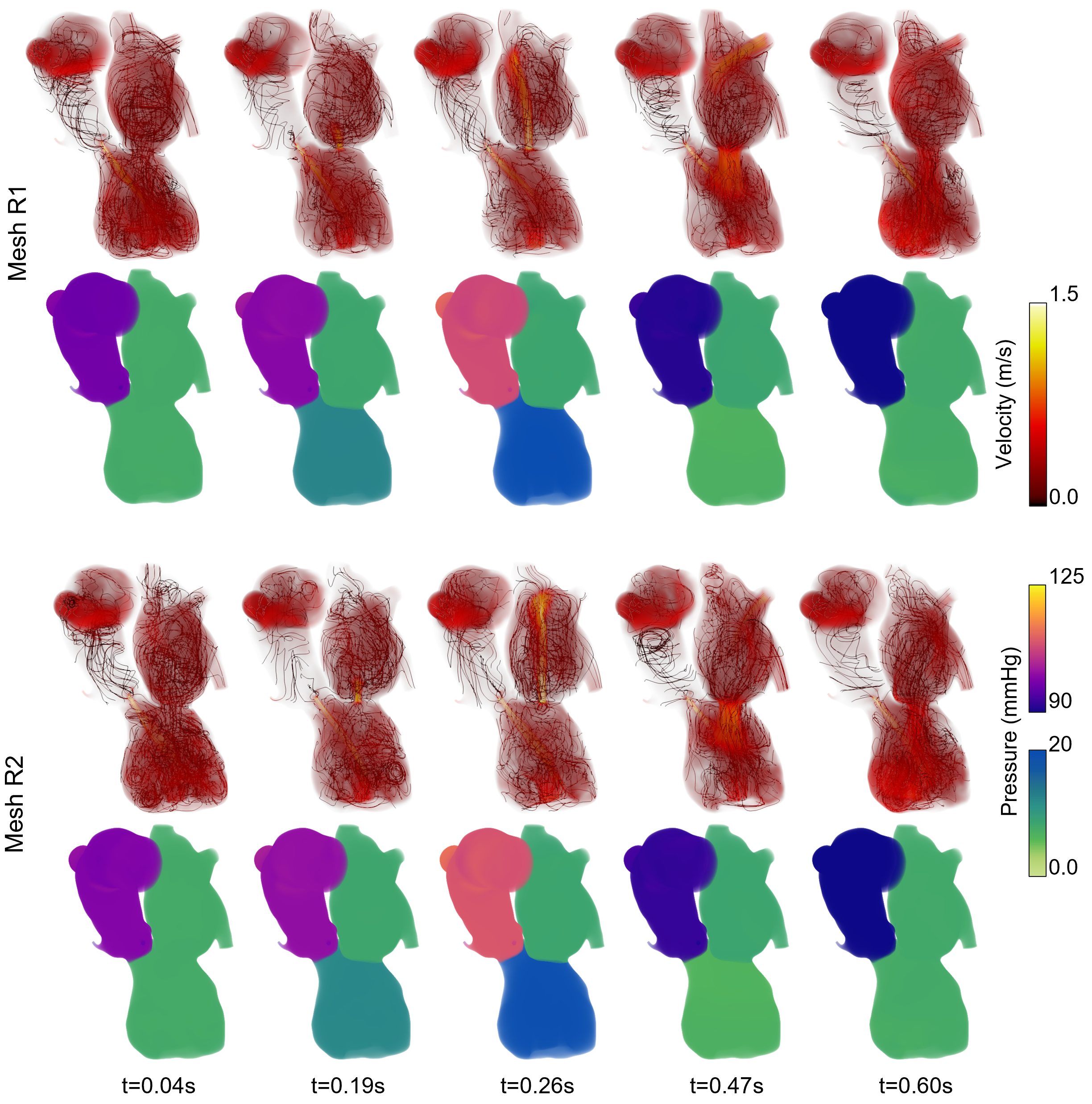}
    \caption{Blood flow and pressure plotted over the cardiac cycle for Patient A with mesh size R1 and R2.}
    \label{fig:mesh3D}
\end{figure}

Figure \ref{fig:mesh3D} compares the blood velocity and pressure fields for mesh sizes R1 and R2 in Patient A. Although slight discrepancies were observed in the timing of jet formation, the overall hemodynamic behavior was consistent between meshes. The supplementary video further demonstrates that the velocity and pressure fields evolve similarly throughout the cardiac cycle, supporting mesh-independent behavior.

\clearpage

\subsection{Lumped Parameter (0D) Model} \label{supp:0d}

The lumped-parameter (0D) model represents the key components of the cardiovascular system, including the systemic system, the coronary circulation, the right ventricle and right atrium, and the pulmonary vasculature. Table \ref{table:0D values} reports the optimized parameters for each patient (A--E). More information about the 0D model, including equations and optimization process, can be found in Bonini et al. \cite{Bonini2025b}.

\begin{longtable}
{|>{\centering\arraybackslash}m{7cm}|>{\centering\arraybackslash}m{1.5cm}|
>{\centering\arraybackslash}m{1.5cm}|
>{\centering\arraybackslash}m{1.5cm}|
>{\centering\arraybackslash}m{1.5cm}|
>{\centering\arraybackslash}m{1.5cm}|}
\hline 
\textbf{Parameter} & \textbf{Pat. A} & \textbf{Pat. B} & \textbf{Pat. C} & \textbf{Pat. D} & \textbf{Pat. E}  \\ 
\hline
\endfirsthead
\hline 
\textbf{Parameter} & \textbf{Pat. A} & \textbf{Pat. B} & \textbf{Pat. C} & \textbf{Pat. D} & \textbf{Pat. E}  \\ 
\hline
\endhead

MV Regurgitant Orifice Area, $A_{o}^{MV}$~(cm$^2$) 
& 0.44 & 0.45 & 0.20 & 0.05 & 0.05\\ \hline
AV Regurgitant Orifice Area, $A_{o}^{AV}$~(cm$^2$) 
& 0.06 & 0.094 & 0.060 & 0.05 & n/a \\ \hline
TV Regurgitant Orifice Area, $A_{o}^{TV}$~(cm$^2$) 
& 0.59 & n/a & 0.20 & 0.40 & 0.30\\ \hline
PV Regurgitant Orifice Area, $A_{o}^{PV}$~(cm$^2$) 
& 0.18 & n/a & n/a & 0.40 & n/a\\ \hline

Systemic Artery Impedance, $Z_{\mathrm{ar}}^{\mathrm{sys}}$~(Pa$\cdot$s/mm$^3$) 
&5.64e-2 & 3.75e-2 & 4.89e-2  & 1.21e-2 & 2.93e-2 \\ \hline
Systemic Artery Capacitance, $C_{\mathrm{ar}}^{\mathrm{sys}}$~(mm$^3$/Pa$\cdot$s)
&  86.53 & 124.71 & 36.71 & 126.98  & 83.39\\ \hline
Systemic Artery Resistance, $R_{\mathrm{ar}}^{\mathrm{sys}}$~(Pa$\cdot$s/mm$^3$) 
& 9.89e-2 & 17.95e-2 & 9.44e-2 & 11.57e-2  & 2.23e-2\\ \hline
Systemic Artery Inductance, $L_{\mathrm{ar}}^{\mathrm{sys}}$~(Pa$\cdot$s$^2$/mm$^3$) 
&0.667e-3& 0.667e-3 & 0.667e-3 & 0.667e-3 & 0.667e-3 \\ \hline
Systemic Vein Capacitance, $C_{\mathrm{ven}}^{\mathrm{sys}}$~(mm$^3$/Pa$\cdot$s)
& 494.33 & 747.39 & 495.17 & 603.28 &376.65 \\ \hline
Systemic Vein Resistance, $R_{\mathrm{ven}}^{\mathrm{sys}}$~(Pa$\cdot$s/mm$^3$) 
& 6.51e-2 & 4.76e-2 & 3.21e-2 & 5.73e-2 & 7.03e-2\\ \hline

Coronary Artery Proximal Impedance, $Z_{\mathrm{cor,p}}^{\mathrm{sys}}$~(Pa$\cdot$s/mm$^3$) 
& 1.71 & 0.88 & 0.97 & 0.796 & 0.65\\ \hline
Coronary Artery Proximal Capacitance, $C_{\mathrm{cor,p}}^{\mathrm{sys}}$~(mm$^3$/Pa$\cdot$s)
& 0.055 & 0.779 & 0.754 & 0.225 & 0.046\\ \hline
Coronary Artery Proximal Resistance, $R_{\mathrm{cor,p}}^{\mathrm{sys}}$~(Pa$\cdot$s/mm$^3$)
&  2.12  & 0.49 & 0.54 & 0.45 &0.36 \\ \hline
Coronary Vein Distal Capacitance, $C_{\mathrm{cor,d}}^{\mathrm{sys}}$~(mm$^3$/Pa$\cdot$s) 
&0.096 & 6.641 & 6.434 & 1.919 &0.393\\ \hline
Coronary Vein Distal Resistance, $R_{\mathrm{cor,d}}^{\mathrm{sys}}$~(Pa$\cdot$s/mm$^3$)  
&6.48& 4.05 & 4.49 & 3.68  &3.01 \\ \hline

RA Minimum Elastance, $E_{\mathrm{min}}^{\mathrm{RA}}$~(Pa/mm$^3$) 
&2.39e-3 & 3.13e-3 & 7.46e-3  & 3.25e-3 & 18.69e-3\\ \hline
RA Maximum Elastance, $E_{\mathrm{max}}^{\mathrm{RA}}$~(Pa/mm$^3$) 
&5.45e-3& 3.84e-3 & 7.86e-3 & 9.45e-3 &18.77e-3 \\ \hline
RV Minimum Elastance, $E_{\mathrm{min}}^{\mathrm{RV}}$~(Pa/mm$^3$)
& 3.76e-3& 2.51-3  & 6.95e-3 & 2.40e-3 & 7.26e-3\\ \hline
RV Maximum Elastance, $E_{\mathrm{max}}^{\mathrm{RV}}$~(Pa/mm$^3$)
& 2.08e-2& 1.95e-2 & 1.28e-2 & 1.47e-2 & 2.03e-2\\ \hline

Pulmonary Artery Capacitance, $C_{\mathrm{ar}}^{\mathrm{pul}}$~(mm$^3$/Pa$\cdot$s)
& 29.98 & 12.82 & 18.58 & 30.17 &4.61 \\ \hline
Pulmonary Artery Resistance, $R_{\mathrm{ar}}^{\mathrm{pul}}$~(Pa$\cdot$s/mm$^3$) 
& 0.96e-2  & 3.67e-2 & 1.83e-2 & 2.25e-2 & 1.76e-2\\ \hline
Pulmonary Vein Capacitance, $C_{\mathrm{ven}}^{\mathrm{pul}}$~(mm$^3$/Pa$\cdot$s)
&  118.35 & 485.28 & 428.80 & 337.76  & 338.90\\ \hline
Pulmonary Vein Resistance, $R_{\mathrm{ven}}^{\mathrm{pul}}$~(Pa$\cdot$s/mm$^3$)  
& 0.11e-3& 0.10e-3 & 3.79e-3  & 0.20e-3 & 0.11e-3\\ 
\hline

\caption{The 0D model parameters fit during optimization. The coronary artery parameters are used for the left ($l$) and right ($r$) coronary arteries.}
\label{table:0D values} \\
\end{longtable}

% ------------------------------

\subsection{Systolic Blood Pressure Analysis pre- to post- TAVR} \label{supp:bp}

We analyzed mean blood pressure (mBP) in eight LVAD patients before and after transcatheter aortic valve replacement (TAVR), performed to treat moderate to severe aortic insufficiency (AI). For each patient, mBP was calculated at each measurement from noninvasive cuff pressure data ($\text{mBP} = \tfrac{1}{3}\text{sBP} + \tfrac{2}{3}\text{dBP}$) and then averaged over the two weeks preceding and the two weeks following the TAVR procedure to obtain the pre- and post-TAVR mBP values, respectively. A two-tailed paired $t$-test revealed no significant difference between pre- and post-TAVR mBP values (\textit{p} = 0.6).

\begin{table}[!htb]
    \renewcommand{\arraystretch}{1.3}
    \centering
    \begin{tabular}{| >{\centering\arraybackslash} m{7cm}| >{\centering\arraybackslash}m{7cm}|}
        \hline
        \textbf{pre-TAVR mBP (mmHg)} & \textbf{post-TAVR mBP (mmHg)} \\ \hline 
        85.6 & 82.4\\ \hline 
        78.3 & 97.7\\ \hline 
        88.0 & 83.7\\ \hline 
        96.3 & 98.6\\ \hline 
        91.0 & 82.0 \\ \hline 
        76.8 & 77.6\\ \hline 
        84.5 & 94.3\\ \hline 
        87.4 & 83.3\\ \hline 
    \end{tabular}
    \caption{Mean systolic blood pressure (mBP) pre- to post- TAVR.}
    \label{supp:table}
\end{table}
\clearpage

\subsection{New 0D Parameters for Simulating Left Heart Valve Repair}
Below are the final calibrated parameters for modeling valve repair as explained in Section \ref{sec:vlvrepair}.

\begin{table}[!htb]
\small
\centering
\renewcommand{\arraystretch}{1.3} % Increases row height by 1.5x
    \begin{tabular}{| >{\centering\arraybackslash} m{4cm}| 
                      >{\centering\arraybackslash}m{3.9cm}|
                      >{\centering\arraybackslash}m{3.9cm}| 
                      >{\centering\arraybackslash}m{3.9cm}|}
        \hline
        & $R_{ar}^{sys}$ (mmHg/mm$^3$) & $Z_{ar}^{sys}$ (mmHg/mm$^3$) & Pump Speed (rpm) \\  \hline
        \textbf{Patient A} & 0.09895 & 0.05642 & 5400 \\ 
        MV Repair & 0.07037 & 0.02621 & 5400 \\
        AV Repair & 0.08281 & 0.05362 & 5400 \\
        MV\&AV Repair & 0.06866 & 0.02622 & 4830 \\ \hline
        \textbf{Patient B} & 0.17952 & 0.03745 & 5900 \\ 
        MV Repair & 0.10873 & 0.01418 & 5900 \\
        AV Repair & 0.11530 & 0.03810 & 5900 \\
        MV\&AV Repair & 0.10873 & 0.01418 & 4780 \\ \hline
        \textbf{Patient C} & 0.09441 & 0.04888 & 5200 \\ 
        MV Repair & 0.06699 & 0.02764 & 5200 \\
        AV Repair & 0.07208 & 0.04689 & 5200 \\
        MV\&AV Repair & 0.05425 & 0.02764 & 5200 \\ \hline
        \textbf{Patient D} & 0.11572 & 0.01206 & 5700 \\ 
        MV Repair & 0.10395 & 0.01127 & 5700 \\
        AV Repair & 0.09853 & 0.01247 & 5700 \\
        MV\&AV Repair & 0.09040 & 0.01131 & 5700 \\ \hline
        \textbf{Patient E} & 0.02226 & 0.02928 & 6000 \\ 
        MV Repair & 0.01345 & 0.02650 & 6000 \\
        AV Repair & n/a & n/a & n/a \\
        MV\&AV Repair & n/a & n/a & n/a \\ \hline
    \end{tabular}
    \caption{Adapted Parameters for simulated  Valve Repair. $R_{ar}^{sys}$, proximal systemic arterial resistance; $Z_{ar}^{sys}$, distal systemic arterial resistance. Patient E did not have AI, and so AV repair was not performed.}
    \label{tab:newparams}
\end{table}

\subsection{Supplemental Results} \label{sec:supp_res}
\begin{figure}[hbt!]
    \centering
    \includegraphics[width=0.9\linewidth]{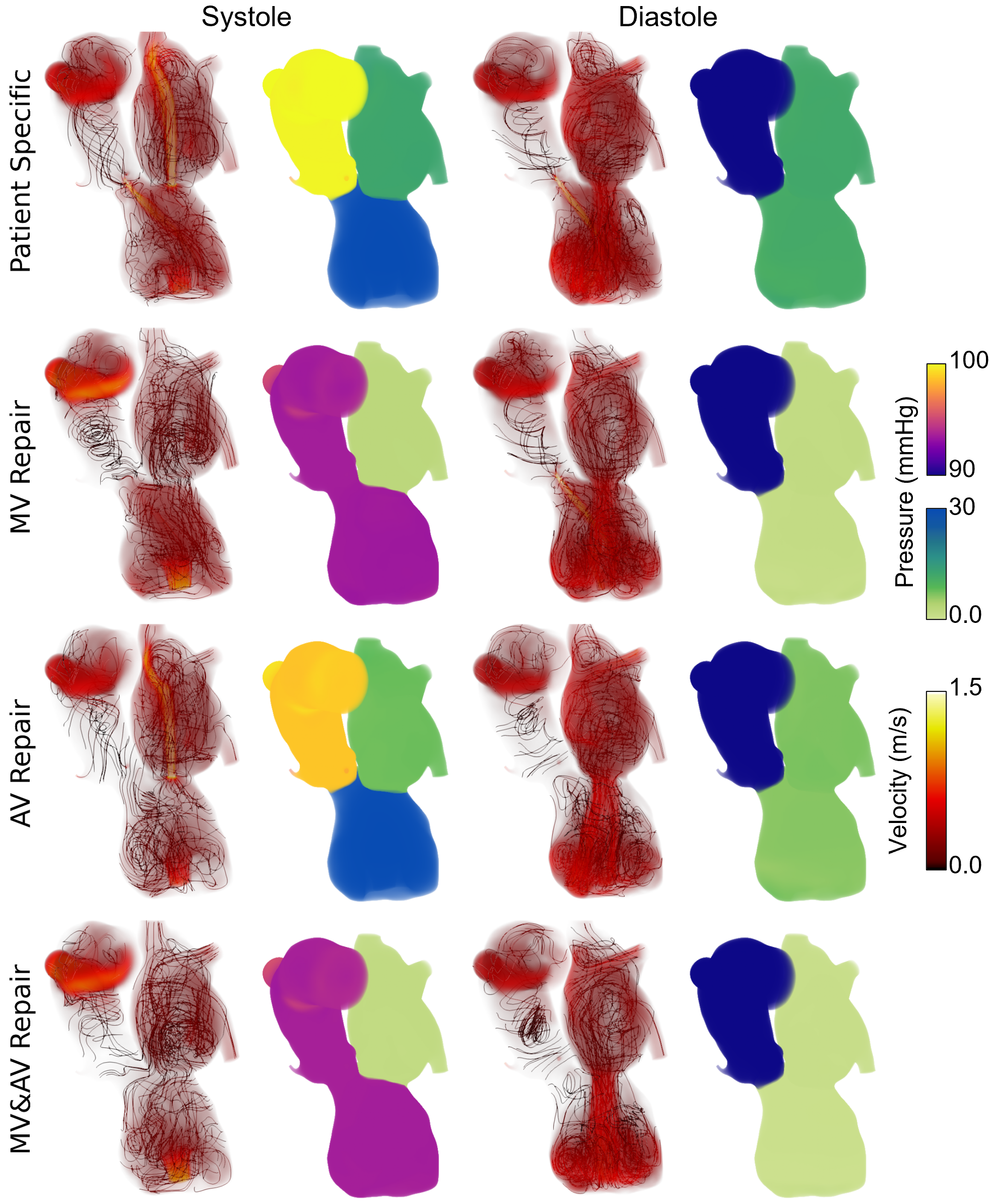}
    \caption{Blood flow and pressure plotted at peak systole and end diastole for Patient A and the results of MV repair, AV repair, and MV\&AV repair.}
    \label{fig:patA}
\end{figure}

\begin{figure}[hbt!]
    \centering
    \includegraphics[width=0.9\linewidth]{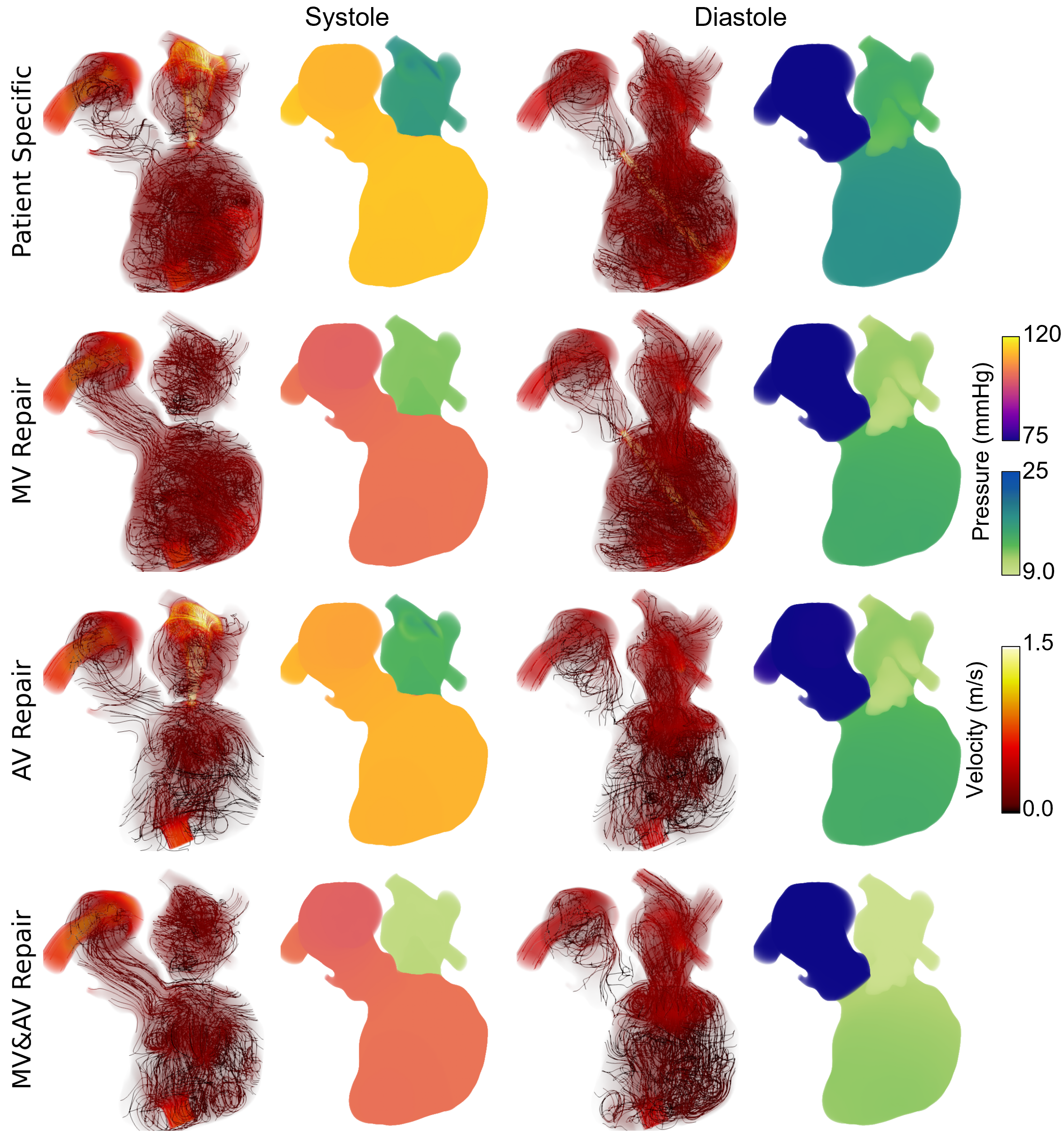}
    \caption{Blood flow and pressure plotted at peak systole and end diastole for Patient C and the results of MV repair, AV repair, and MV\&AV repair.}
    \label{fig:patC}
\end{figure}

\paragraph{Supplemental Hemodynamic Results}
Figures \ref{fig:patA} and \ref{fig:patC} show the blood velocity and pressure fields at peak systole and end-diastole for Patients A and C.

\clearpage
\begin{figure}[hbt!]
    \centering
    \includegraphics[width=0.88\linewidth]{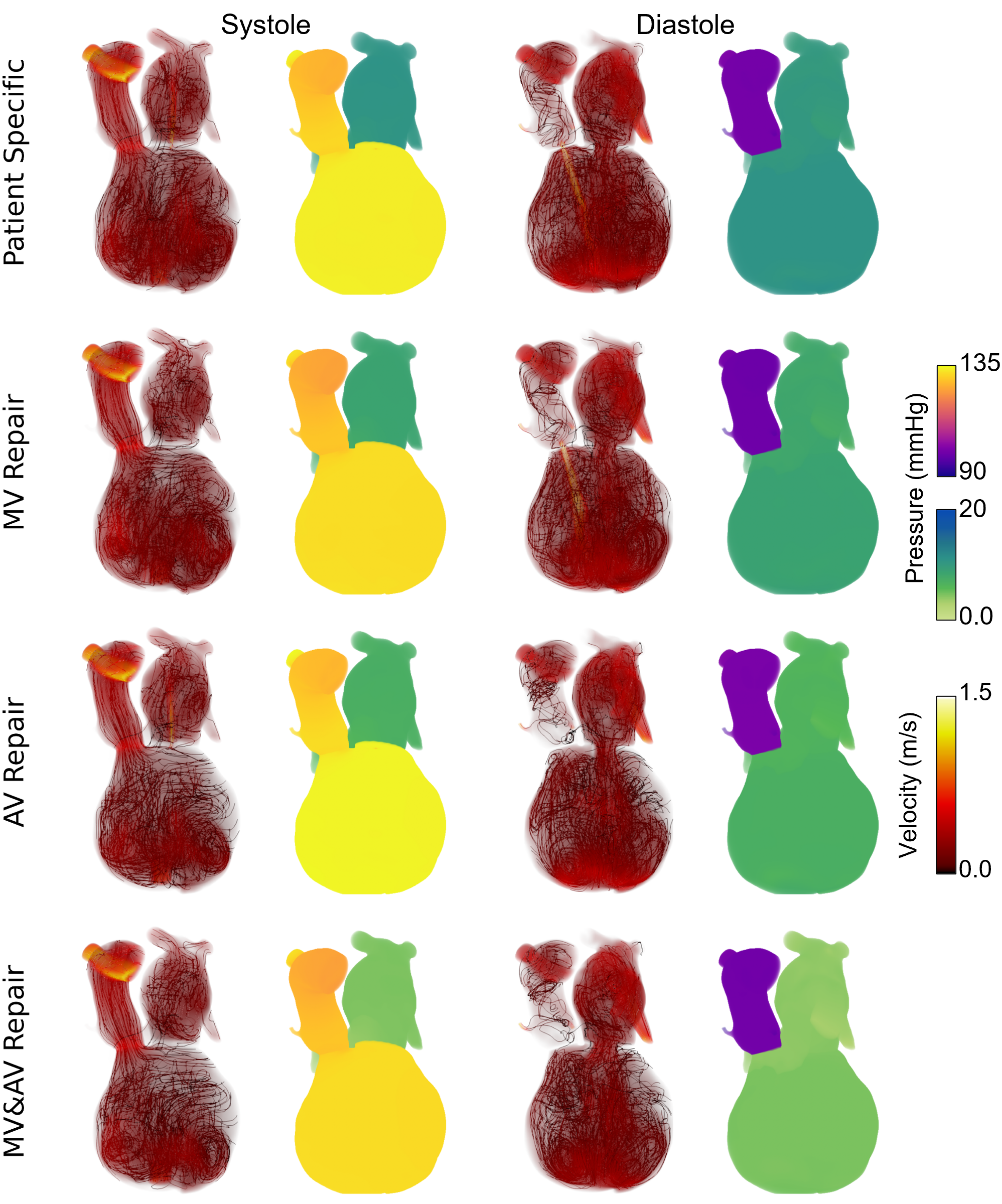}
    \caption{Blood flow and pressure plotted at peak systole and end diastole for Patient D and the results of MV repair, AV repair, and MV\&AV repair.}
    \label{fig:patD}
\end{figure}
\begin{figure}[hbt!]
    \centering
    \includegraphics[width=0.8\linewidth]{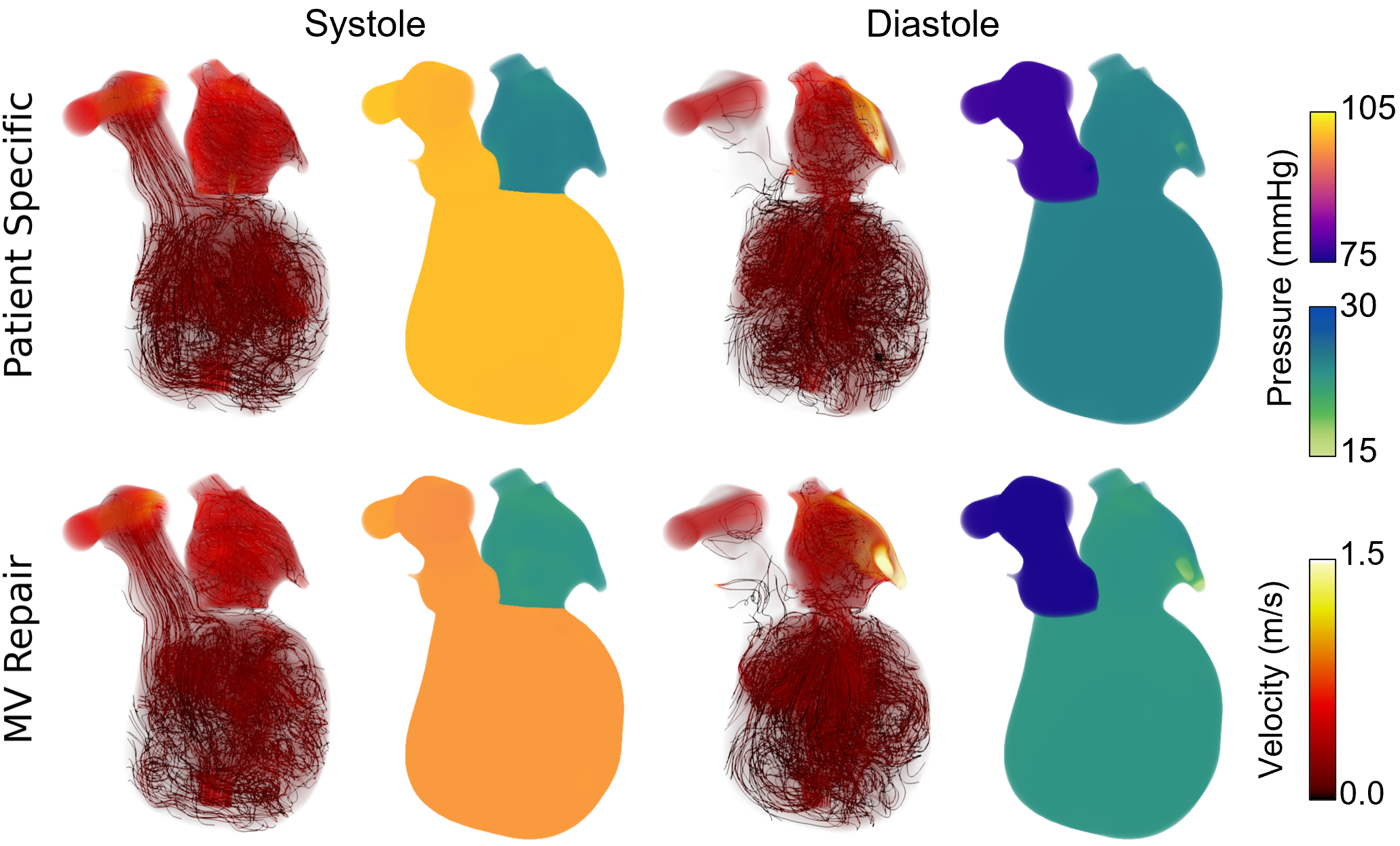}
    \caption{Blood flow and pressure plotted at peak systole and end diastole for Patient E and the results of MV repair.}
    \label{fig:patE}
\end{figure}

The percentage of the cardiac cycle with forward AV flow varied across patients and interventions. For Patient A, values were 0\%, 5.9\%, 0\%, and 6.4\% for the patient-specific, MV repair, AV repair, and MV\&AV repair cases, respectively. For Patient B, corresponding values were 4.0\%, 9.6\%, 3.8\%, and 12.4\%, and for Patient C, 1.7\%, 15.7\%, 1.7\%, and 15.6\%. Patients D and E exhibited higher baseline forward flow with smaller relative changes: Patient D showed 21.1\%, 22.1\%, 21.2\%, and 22.2\%, while Patient E showed 27.3\% and 28.3\%.
%Concurrently, tricuspid valve regurgitant fraction (TVRF) decreased, indicating secondary relief of RV volume overload once left‐sided pressures were normalized.

\begin{figure}[htb!]
    \centering
    \includegraphics[width = 0.7\textwidth]{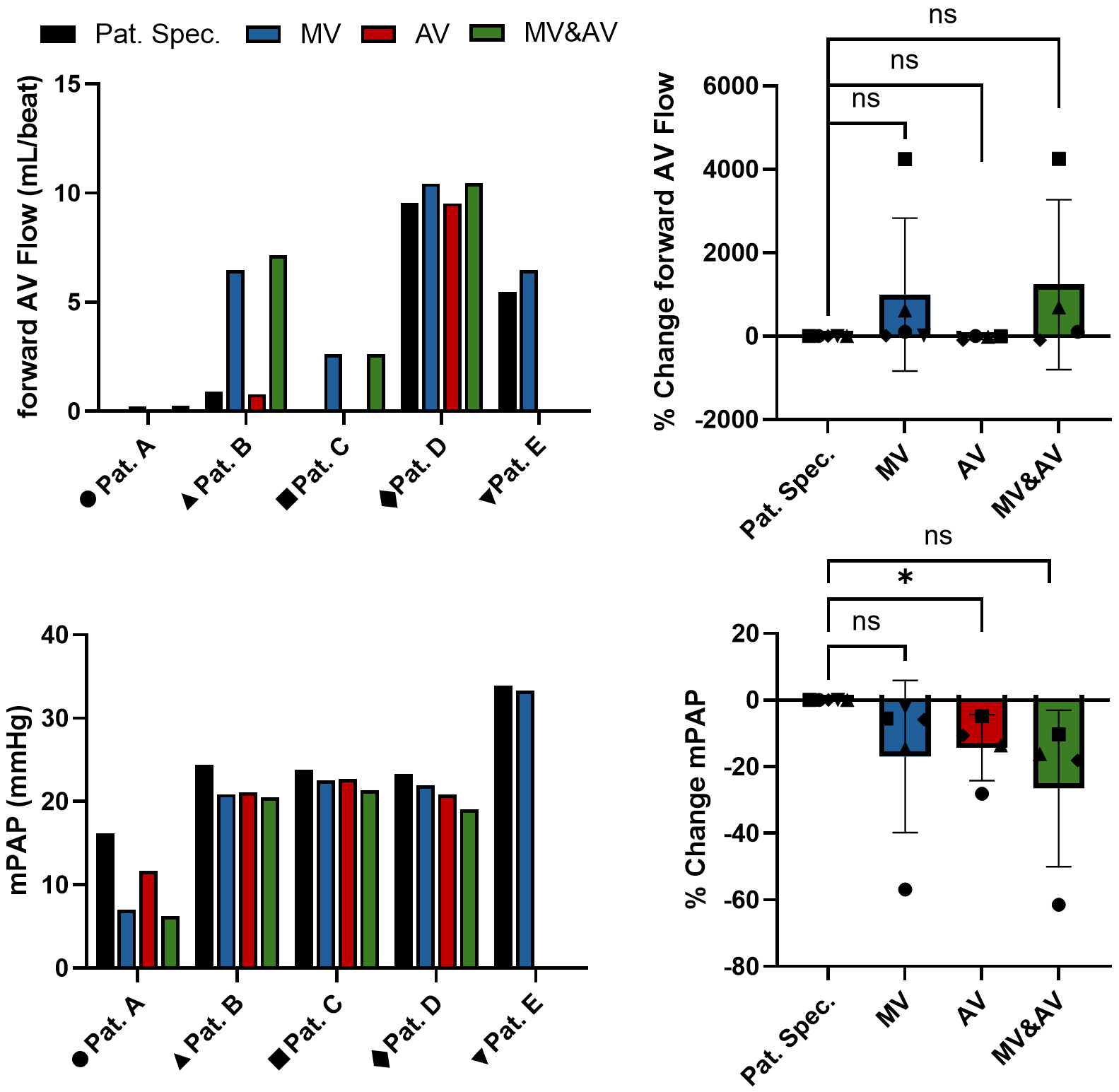}
    \caption{The values for each metric are plotted on the left grouped by patient. Black bars are the patient-specific results, blue bars are the results for \greentxt{simulated} MV repair, red bars for \greentxt{simulated} AV repair, and green bars for \greentxt{simulated} AV\&MV repair. The plots on the right are showing the average percent increase in each value ($y$) compared to the patient-specific case ($y_{ps}$) for all patients, $(y-y_{ps})/y_{ps} \cdot 100$. And the significance of the average percent increase with respect to the patient-specific cases. The data is plotted with circles for Patient A, triangles for Patient B, squares for Patient C, diamonds for Patient D, and upside down triangles for Patient E.}
    \label{fig:bars_supp}
\end{figure}

\begin{figure}[!htb]
    \centering
    \includegraphics[width=0.8\textwidth]{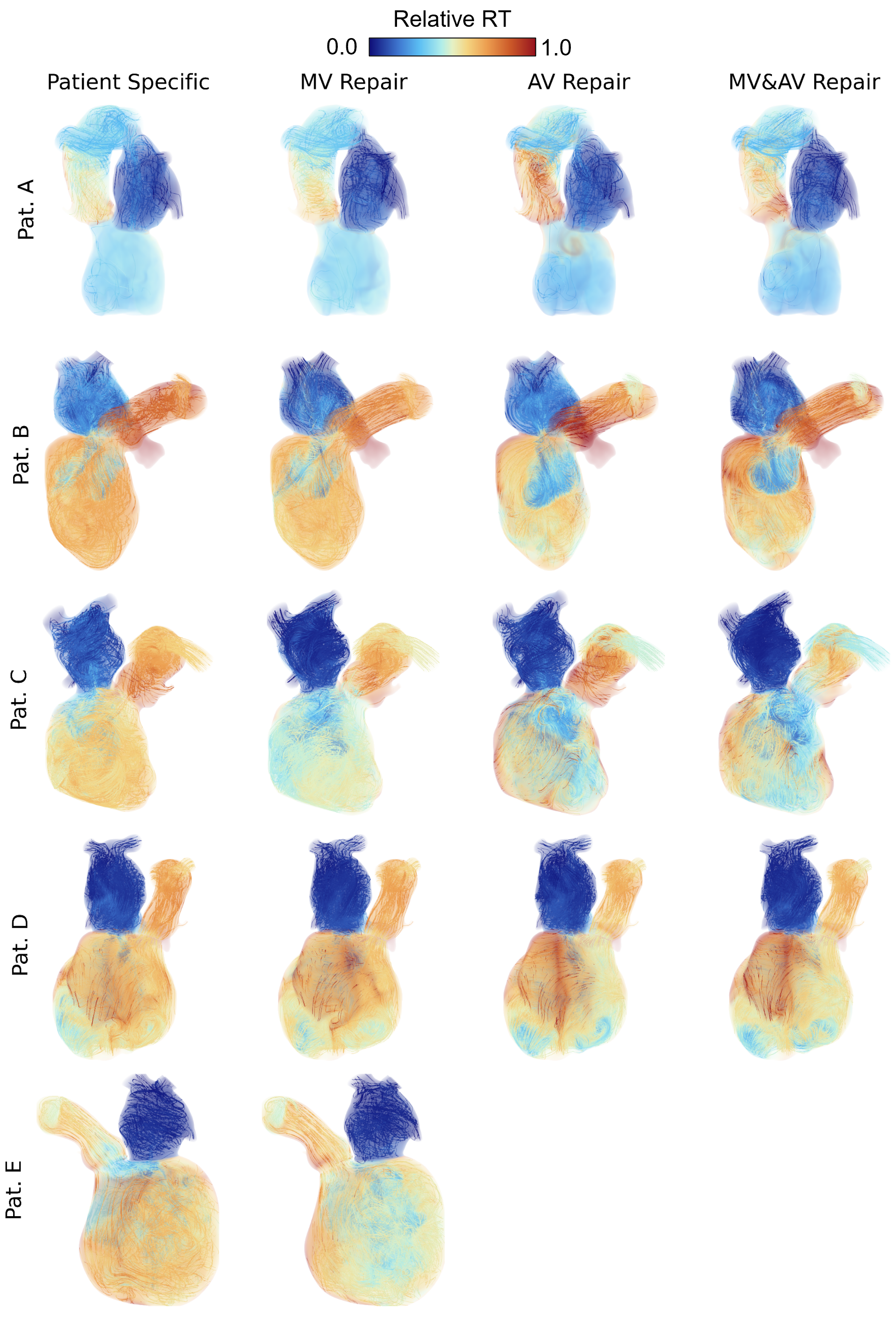} 
    \caption{Relative residence time for Patient's A--E at peak systole for all cases of valve repair. Relative residence time is visualized using volume rendering, with streamlines depicting flow direction and colored according to local relative residence time.}
    \label{fig:rtlh} 
\end{figure}

\begin{figure}[!htb]
    \centering
    \includegraphics[width=0.8\textwidth]{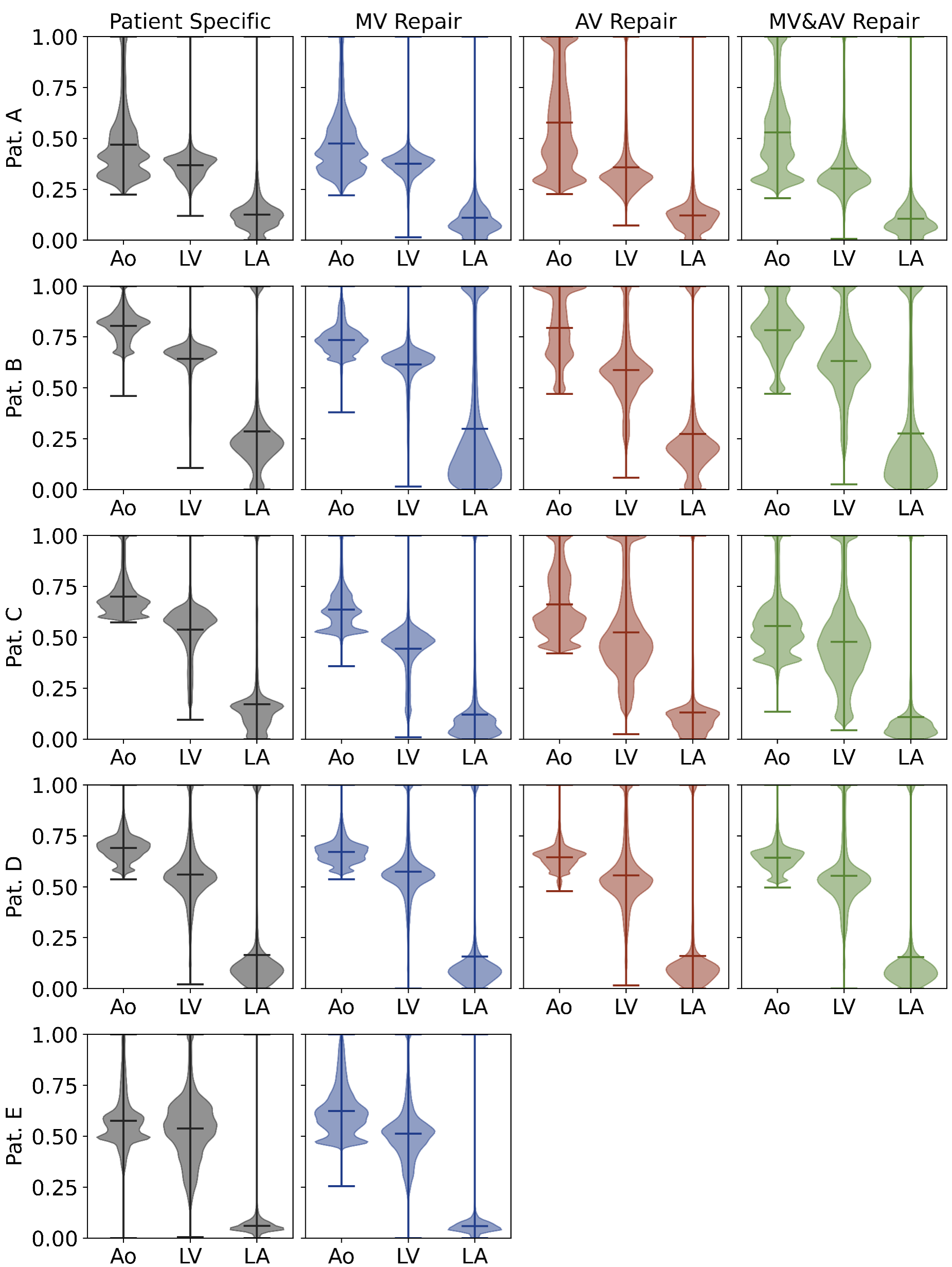} 
    \caption{Violin plots showing the relative residence time at peak systole for the aorta ($\Omega_{Aorta}$), LV  ($\Omega_{LV}$), and  ($\Omega_{LA}$).}
    \label{fig:rtvio} 
\end{figure}

\paragraph{Relative Residence Time} \label{supp:rt}
Across all patients, our simulations demonstrated prolonged relative residence times (rRT) within the left ventricle and aorta, particularly in regions of recirculating flow within the aortic sinuses \ref{fig:rtlh}. Figure \ref{fig:rtvio} shows the distribution of the relative residence time for each domain ($\Omega_{Aorta},~\Omega_{LV},$ and $\Omega_{LA}$). 
Patient~A showed the lowest mean relative RT values (Ao/LV/LA: 0.47/0.36/0.11), 
Patient~B exhibited the highest overall mean rRT of (0.80/0.64/0.29), 
Patient~C's mean rRT were (0.70/0.54/0.17), Patient D were (0.59/0.56/0.165), and Patient E's were (0.576/0.538/0.06), indicating variability in baseline washout efficiency across the cohort. 
For Patients~B and~C MV repair led to a significant increase in forward aortic flow, which corresponded to reductions in aortic root rRT of 8.6\% and 9.1\%, respectively. In contrast, Patient's~A, D, and E exhibited minimal changes in forward flow, producing little change in aortic rRT. 
Aortic valve (AV) repair increased the mean rRT in the aorta in patient A (23.0\%) and decreased aortic rRT in patients B (1.2\%), C (5.4\%), and D (6.67\%). For all 4 patients (A--D), the LV rRT decreased after AV repair by 0.9–-8.6\%. 
Combined MV\&AV repair produced the largest and most consistent reductions across compartments, particularly in the LA (up to $\sim$37\%) and LV (up to $\sim$11\%), indicating synergistic improvements in washout when both valves are addressed. While mean aortic RT responses remained more variable, the combined intervention generally yielded greater overall reductions compared to isolated repairs, supporting a cumulative effect of dual-valve treatment on global flow efficiency.

\clearpage

\begin{figure}[!htb]
    \centering
    \includegraphics[width=0.8\textwidth]{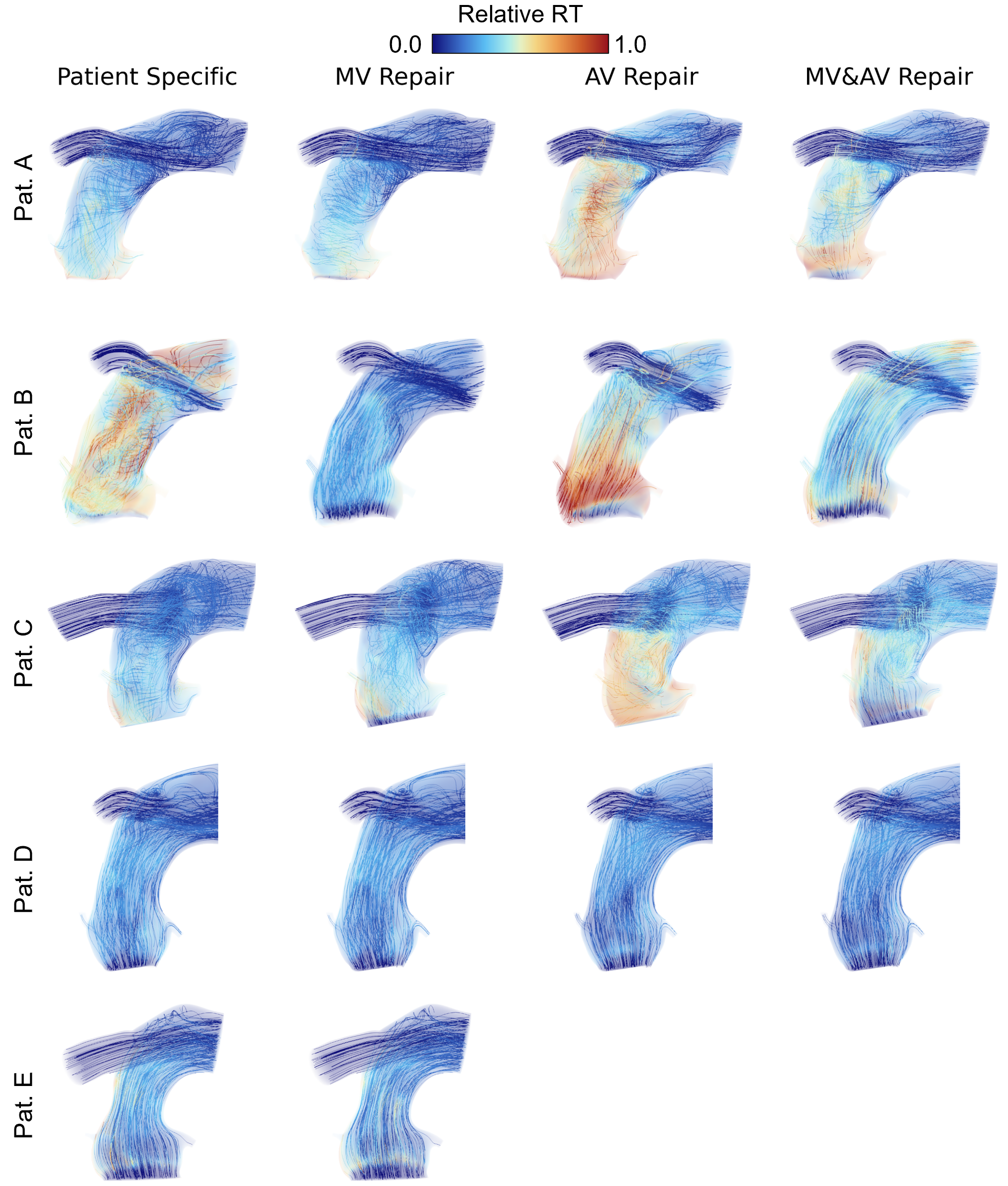} 
    \caption{Isolated Aortic relative residence time for Patient's A--E at peak systole for all cases of valve repair. Relative residence time is visualized using volume rendering, with streamlines depicting flow direction and colored according to local relative residence time.}
    \label{fig:rtao} 
\end{figure}

\begin{figure}[!htb]
    \centering
    \includegraphics[width=0.8\textwidth]{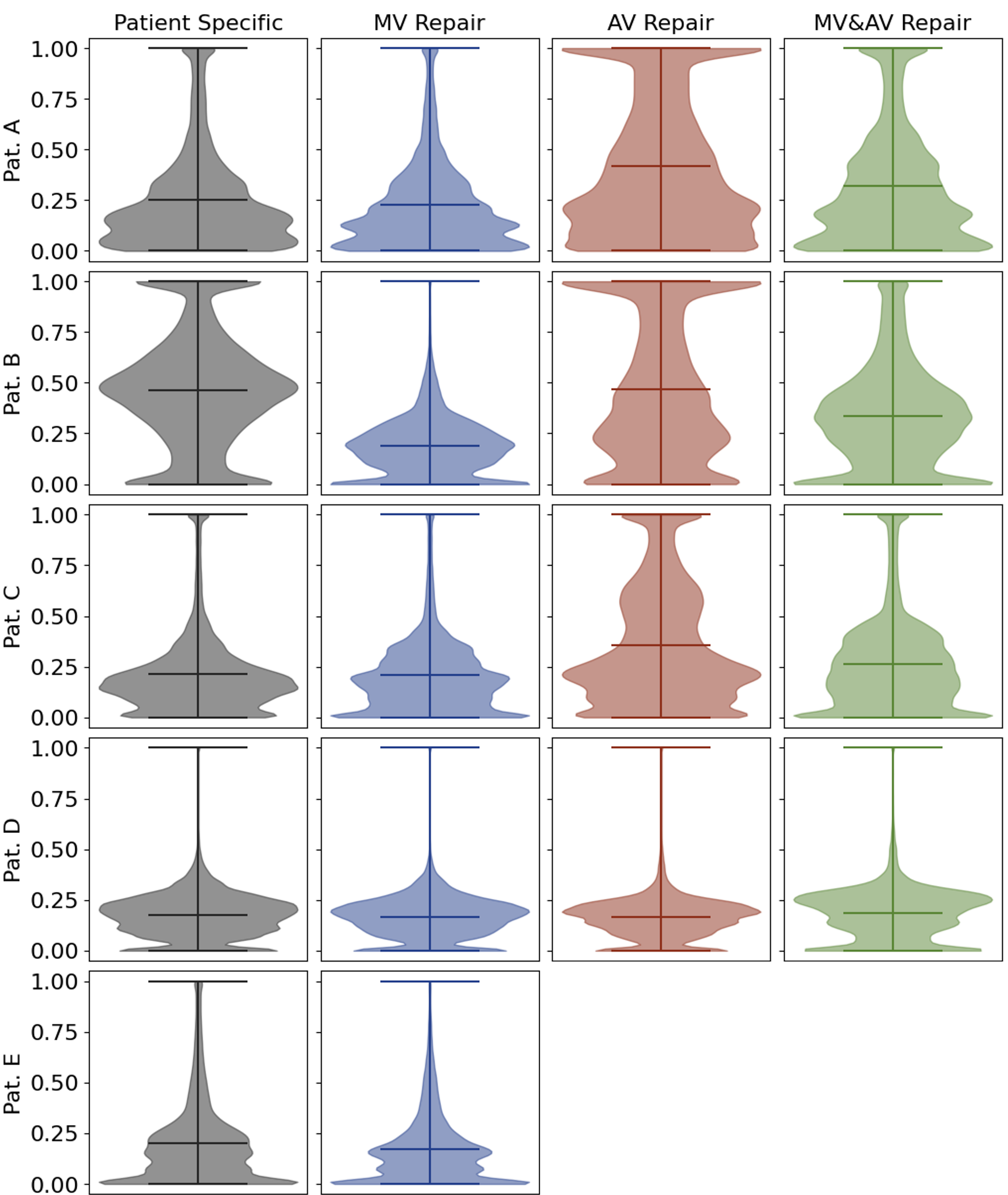} 
    \caption{Violin plots showing the relative residence time for the aortic domain with Dirichlet boundary conditions ($\bv=0$) for flow entering the model.}
    \label{fig:rtaovio} 
\end{figure}

Relative residence time (rRT) results for the isolated aortic domain (Equations \ref{eq:rtao}), is shown in Figure \ref{fig:rtao}. 
Here we see that the aortic sinuses showed the highest rRT. 
The distribution in relative residence is shown in Figure \ref{fig:rtaovio}. 
Compared to the patient-specific case, \greentxt{simulated} mitral valve repair showed a decrease in the mean rRT of 11.5\%, 59.3\%, 2.49\%, 21.7\%, 19.3\% for patients A--E, respectively. 
\greentxt{Simulated} aortic valve repair caused an increase of 65.5\%, 0.34\%, and 68.7\% for patients A--C. There was a decrease of 22.5\% for patient D. 
Concurrent mitral and aortic valve repair led to a 25.6\% increase in Pat. A, 23.9\% decrease in Pat. B, a 23.5\% increase in Pat. C, and a 11.8\% decrease in Pat. D.

\newpage

\bibliographystyle{elsarticle-num-names} 
\bibliography{ref}

\end{document}